\documentclass[preprint,12pt,authoryear]{elsarticle}

\makeatletter
\def\ps@pprintTitle{%
   \let\@oddhead\@empty
   \let\@evenhead\@empty
   \def\@oddfoot{\reset@font\hfil\thepage\hfil}
   \let\@evenfoot\@oddfoot
}
\makeatother

\usepackage{latexsym,bm}
\usepackage{amssymb,amsmath,amsfonts}
\usepackage[usenames]{color}
\usepackage{graphicx}
\usepackage{url}
\usepackage{cancel}
\usepackage{algorithm}
\usepackage{algpseudocode}
\usepackage{tabularx}

\usepackage[markup=default]{changes}
\usepackage{booktabs}
\usepackage{subcaption}
\usepackage{xr}
\newcommand{\change}[1]{\textcolor{black}{#1}}

\makeatletter
\newcommand*{\rom}[1]{\expandafter\@slowromancap\romannumeral #1@}
\newcommand{\multiline}[1]{%
  \begin{tabularx}{\dimexpr\linewidth-\ALG@thistlm}[t]{@{}X@{}}
    #1
  \end{tabularx}
}
\makeatother

\begin{document}

\begin{frontmatter}

\title{\change{Assessing} Spatial Stationarity and Segmenting Spatial Processes into Stationary Components}

\author[nsysu]{ShengLi Tzeng}
\author[purdue]{Bo-Yu Chen}
\author[as]{Hsin-Cheng Huang\corref{cor1}}
\ead{hchuang@stat.sinica.edu.tw}

\address[nsysu]{Department of Applied Mathematics, National Sun Yat-Sen University, Taiwan.}
\address[purdue]{Department of Statistics, Purdue University, USA.}
\address[as]{Institute of Statistical Science, Academia Sinica, Taiwan.}

\cortext[cor1]{Corresponding author}

\date{}

\begin{abstract}
\baselineskip=24pt
\change{In this research, we propose a novel technique for visualizing nonstationarity in geostatistics,
particularly when confronted with a single realization of data at irregularly spaced locations.
Our method hinges on formulating a statistic that tracks a stable microergodic parameter of the exponential covariance function,
allowing us to address the intricate challenges of nonstationary processes that lack repeated measurements.
We implement the fused lasso technique to elucidate nonstationary patterns at various resolutions.
For prediction purposes, we segment the spatial domain into stationary sub-regions via Voronoi tessellations.
Additionally, we devise a robust test for stationarity based on contrasting the sample means of our proposed statistics between two selected Voronoi subregions.
The effectiveness of our method is demonstrated through simulation studies and its application to a precipitation dataset in Colorado.}
\bigskip
\end{abstract}


\end{frontmatter}

\baselineskip=24pt

\newpage
\section{Introduction}

Consider a spatial process $\{y(\bm{s}):\bm{s}\in D\}$ of interest defined on a region $D\subset\mathbb{R}^2$.
Suppose that we observe data $\bm{z}\equiv(z(\bm{s}_1),\dots,z(\bm{s}_n))'$ at $n$ spatial locations,
which may be irregularly spaced, according to the measurement equation:
\begin{equation}
z(\bm{s}_i)=y(\bm{s}_i)+\change{e}(\bm{s}_i);\quad i =1,\dots,n,
\label{eq:data}
\end{equation}

\noindent where $\change{e}(\bm{s}_1),\dots,\change{e}(\bm{s}_n)\sim N(0,\change{\tau^2})$ are white-noise variables,
representing measurement errors.
A major problem in geostatistics, called kriging, is to predict $y(\bm{s}_0)$ at any location $\bm{s}_0\in D$ based on $\bm{z}$.
For simplicity, we assume that the mean function of the process $y(\cdot)$ is known and, without loss of generality, zero.
Then for a given covariance function of $y(\cdot)$, \change{the ordinary-kriging} predictor of $y(\bm{s}_0)$ is
\begin{equation}
\change{\hat{y}(\bm{s}_0)=\left(\bm{c}+\frac{1-\bm{c}'\bm{\Sigma}^{-1}\textbf{1}}{\textbf{1}'\bm{\Sigma}^{-1}\textbf{1}}\textbf{1}\right)'\bm{\Sigma}^{-1}\bm{Z},}
\label{eq:ok}
\end{equation}

\noindent \change{where $\bm{c}\equiv\mathrm{cov}(\bm{z},y(\bm{s}_0))$, $\bm{\Sigma}\equiv\mathrm{var}(\bm{z})$, and $\textbf{1}=(1,\dots,1)'$.}

\begin{figure}\centering
\begin{tabular}{cc}
\includegraphics[scale=0.3,trim={1cm 4.5cm 0.5cm 3.5cm},clip]{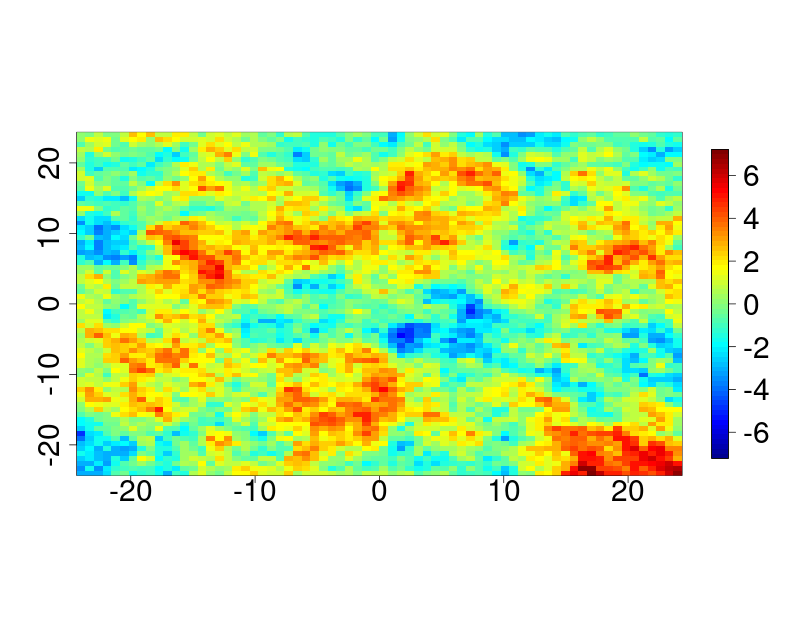} &
\includegraphics[scale=0.3,trim={1cm 4.5cm 0.5cm 3.5cm},clip]{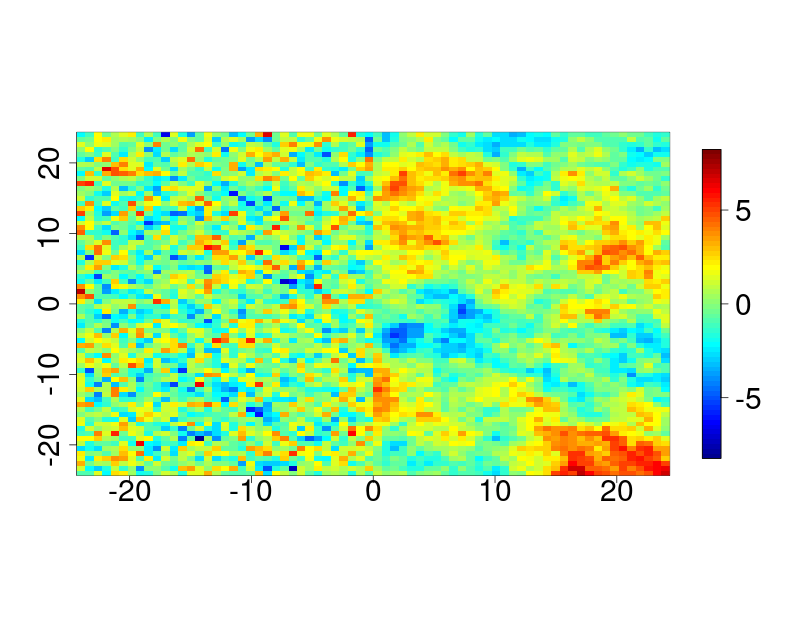} \\
(a1) & (a2)\\
\includegraphics[scale=0.3,trim={1cm 4.5cm 0.5cm 3.5cm},clip]{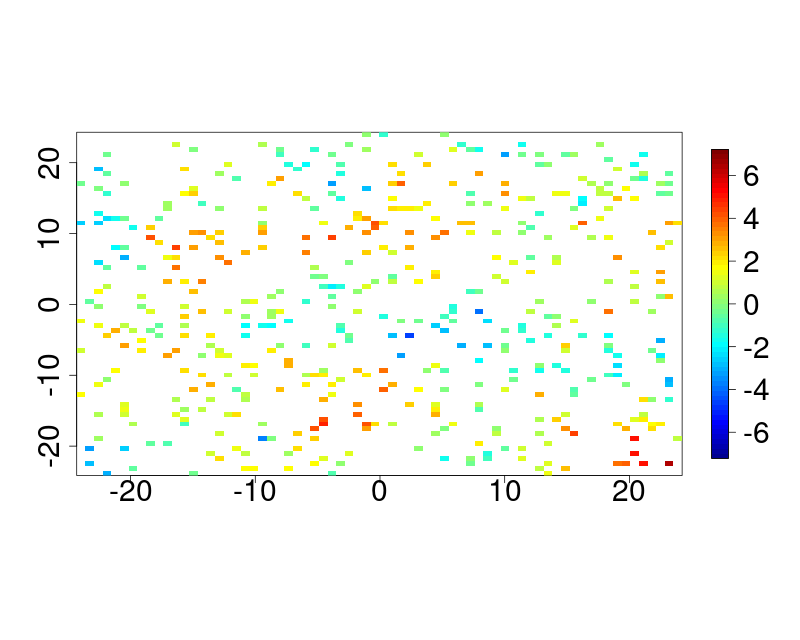} &
\includegraphics[scale=0.3,trim={1cm 4.5cm 0.5cm 3.5cm},clip]{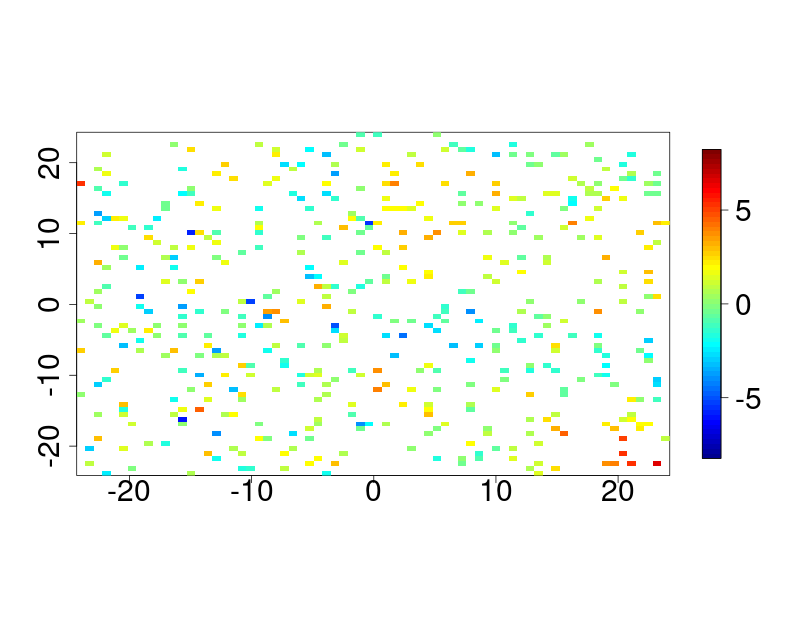} \\
(b1) & (b2) \\
\includegraphics[scale=0.3,trim={1cm 4.5cm 0.5cm 3.5cm},clip]{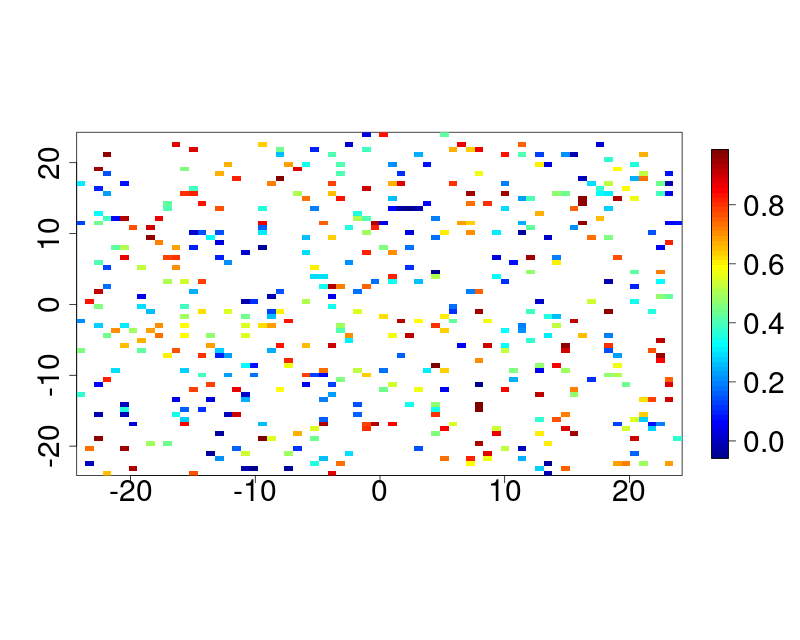} &
\includegraphics[scale=0.3,trim={1cm 4.5cm 0.5cm 3.5cm},clip]{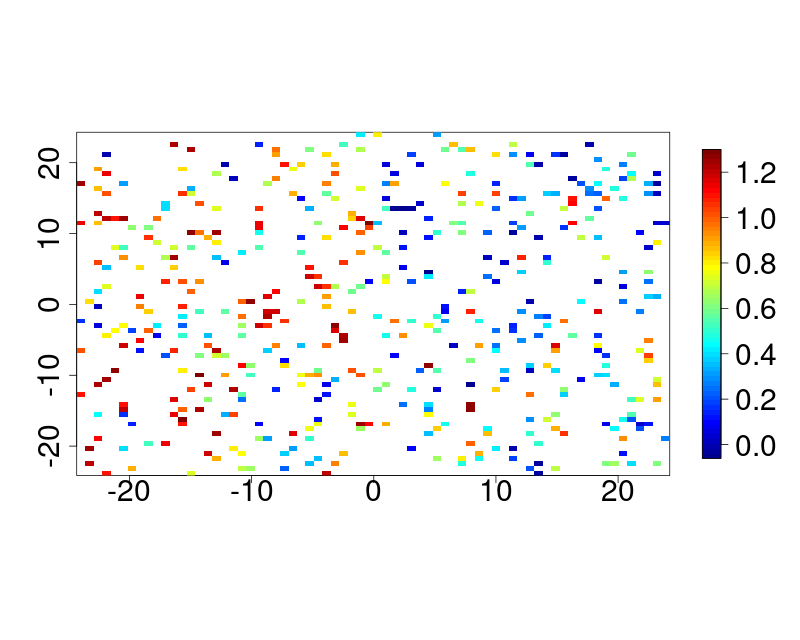} \\
(c1) & (c2) \\
\end{tabular}
\caption{\change{(a1) A zero-mean stationary process; (a2) A zero-mean nonstationary process;
  (b1) Data sampled from the process in (a1) at 400 locations using simple random sampling;
  (b2) Data sampled from the process in (a2) at 400 locations using simple random sampling;
  (c1) Local spatial indices based on the data in (b1);
  (c2) Local spatial indices based on the data in (b2).
}}
\label{fig:motivation}
\end{figure}

Given a realization noisy data $\bm{z}$ at $n$ locations, \change{it is typical to assume that the covariance function of $y(\cdot)$ is stationary.}
A commonly used stationary covariance model is the isotropic Mat\'{e}rn family (Mat\'{e}rn, 1986) given by
\begin{equation}
\change{\mathrm{cov}(y(\bm{s}),y(\bm{s}+\bm{u}))}
=\frac{\change{\sigma^2}}{2^{\nu-1}\Gamma(\nu)}\bigg(\frac{\sqrt{2\nu}}{\alpha}\|\bm{u}\|\bigg)^{\nu}\mathcal{K}_{\nu}
\bigg(\frac{\sqrt{2\nu}}{\alpha}\|\bm{u}\|\bigg);\quad\bm{s},\,\bm{s}+\bm{u}\in\mathbb{R}^2,
\label{eq:Matern}
\end{equation}

\noindent where $\mathcal{K}_\nu(\cdot)$ is the modified Bessel function of the second kind of order $\nu$,
$\change{\sigma^2}$ is a variance parameter,
and $\bm{\theta}\,\equiv\,(\alpha,\nu)'$ consists of a scale parameter $\alpha$ and a smoothness parameter $\nu$.
\change{It's important to emphasize that spatial covariance functions don't always exhibit stationarity.
Sometimes, they can be markedly influenced by local conditions and topographical variations,
leading to substantial deviations from stationarity.
Visualizing the nonstationary attributes from a single dataset presents a challenge.
For illustration, Figure~\ref{fig:motivation}(a1) depicts a zero-mean stationary process with a Matérn covariance function. In contrast, Figure~\ref{fig:motivation}(a2) presents a zero-mean piecewise stationary process characterized by two distinct Mat\'{e}rn covariance functions.
While the nonstationarity in Figure~\ref{fig:motivation}(a2) is apparent, determining which one among Figures~\ref{fig:motivation}(b1) and \ref{fig:motivation}(b2) (consisting of 400 random samples from the processes in Figures~\ref{fig:motivation}(a1) and \ref{fig:motivation}(a2) respectively) exhibits nonstationarity is not straightforward.}

\change{Several approaches have been proposed for testing spatial stationarity.
Fuentes (2005) pioneered a frequency domain test for spatial samples on a regular grid.
Jun and Genton (2012) proposed a test that partitions the spatial domain into two non-intersecting fields for irregularly spaced data.
More recently, Bandyopadhyay and Rao (2017) unveiled a test leveraging the Fourier transform in the frequency domain, catering to irregularly spaced data. 
In addition, local indicators of spatial autocorrelation (LISA) have been proposed by Anselin (1995) for lattice data.
To our knowledge, there seems to be an absence of spatial dependence indices crafted explicitly for irregularly sampled data within geostatistics.}

\change{In this study, we introduce a local statistic designed to highlight nonstationary characteristics within geostatistical datasets.
This is achieved by employing a robust local estimation of a microergodic parameter inherent to the exponential covariance model.
We leverage the fused lasso methodology to illuminate nonstationary patterns across varying resolutions.
To enhance the accuracy of spatial predictions using stationary models, we segment the spatial domain into homogenous sub-regions utilizing Voronoi tessellations.
A rigorous test for spatial stationarity is established by comparing the sample means of the estimated microergodic parameters between two Voronoi subregions.
If the stationarity is violated, we further partition $D$  into  $K$ components $\{D_1,\dots,D_K\}$
such that each process $\{y(\bm{s}):\bm{s}\in D_k\}$ is stationary, for $k=1,\dots,K$.
It's worth mentioning that both Guinness and Fuentes (2015) and Muyskens \textit{et al}.~(2022) have crafted techniques to segregate domain $D$ into stationary subregions.
However, the approach by Guinness and Fuentes requires data on a regular grid,
while the method by Muyskens \textit{et al}.~(2022) necessitates a regular grid shape for their base partition due to the algorithm's reliance on circulant embedding.
It's important to note that the choice of grid resolution influences segmentation and increases computational demands when selecting among different resolutions.}

The rest of this paper is organized as follows.
In Section 2, we develop a statistic to monitor spatial heterogeneity
\change{at multiple resolutions.
The statistic is designed to estimate a microergodic parameter of the exponential variogram with data that may be irregularly spaced.}
We then provide an approach to partition the domain $D$ into $K$ homogeneous subregions using Voronoi tessellations (Voronoi, 1908).
Section 3 gives the proposed test for stationarity using a $t$-type statistic based on the Voronoi \change{subregions} obtained from Section 2 with $K=2$.
Some simulation results are given in Section 4. An application to a precipitation dataset in Colorado is provided in Section 5.
Finally,  Section 6 concludes with a summary.

\section{Segmenting spatial processes into stationary components}

\subsection{A statistic for local spatial dependence}

First, we construct a statistic to monitor the local spatial dependence of $y(\cdot)$ around $\bm{s}_i$, for $i=1,\dots,n$.
To find local structure around $\bm{s}_i$, we consider a neighborhood set of $\bm{s}_i$:
\begin{equation}
N_i\equiv\{j:\|\bm{s}_j-\bm{s}_i\|\leq r,\,j\neq i\};\quad i=1,\dots,n,
\label{eq:neighborhood}
\end{equation}

\noindent where $\|\cdot\|$ is the Euclidean distance and $r>0$ is an appropriate radius.
If $y(\cdot)$ is an isotropic stationary process around $\bm{s}_i$,
then its variogram at distance $h$ is
\[
2\gamma_{y,i}(h)\equiv\mathrm{E}(y(\bm{s})-y(\bm{s}_i))^2,\mbox{ for }h=\|\bm{s}-\bm{s}_i\|\leq r;\quad i=1,\dots,n.
\]
It follows that for \change{$j\in N_i$},
\[
\change{2\gamma_{z,i}(\|\bm{s}_j-\bm{s}_i\|)\equiv\mathrm{E}\big{(}(z(\bm{s}_j)-z(\bm{s}_i))^2\big{)}=2\gamma_y(\|\bm{s}_j-\bm{s}_i\|)+2\delta_{ij}\tau^2};\quad i=1,\dots,n,
\]
where $\delta_{ij}\equiv1$ if $i=j$; $0$ otherwise.
To reflect the local behavior, it is desirable to consider $N_i$'s with a small $r$.
To obtain a statistic that is robust to outliers,
we utilize a squared-root transform and apply the following approximation formula (Cressie and Hawkins, 1980):
\begin{equation}
\frac{|z(\bm{s}_j)-z(\bm{s}_i)|^{1/2}}{\{2\gamma_{z,i}(\|\bm{s}_j-\bm{s}_i\|)\}^{1/4}}\approx \mathcal{N}\big(2^{1/4}\pi^{-1/2}\Gamma(3/4),\,2^{1/2}\{\pi^{-1/2}-\pi^{-1}\Gamma(3/4)^2\}\big);
\quad\bm{s}_j\in N_i.
\label{eq:CH}
\end{equation}

\change{We first assume that $\tau^2$ is known, and consider a local exponential semi-variogram:}
\[
\change{\gamma_{y,i}(h)=\sigma^2_i(1-\exp(h/\alpha_i));\quad 0\geq h,\,i=1,\dots,n,}
\]
\change{parametrized by variance $\sigma_i^2$ and range parameter $\alpha_i$; $i=1,\dots,n$.
However, it is well known that both $\sigma_i^2$ and $\alpha_i$ are unidentifiable under the infill asymptotic framework (Zhang, 2004).
Instead, we focus on their ratio, $\sigma_i^2/\alpha_i$, a microergodic parameter that can be consistently estimated.
Applying a Taylor expansion to $(2\gamma_{z,i}(h))^{1/4}$ at $h=0$, we obtain}
\begin{align*}
  \gamma_{z,i}(h)^{1/4}
=&~ \change{\big\{\sigma^2_i(1-\exp(-h/\alpha_i))+\tau^2\big\}^{1/4}}\nonumber\\
=&~ \change{\tau^{1/2}+\tau^{-3/2}\sigma^2_i h/(4\alpha_i)+O(h^2)};\quad i=1,\dots,n.
\label{eq:Taylor}
\end{align*}

\noindent Substituting $\change{\tau^{1/2}+\tau^{-3/2}\sigma^2_i\|\bm{s}_j-\bm{s}_i\|/(4\alpha_i)}$ above
for $\gamma_{z,i}(\|\bm{s}_j-\bm{s}_i\|)^{1/4}$ in \eqref{eq:CH} leads to
\begin{equation}
\change{\mathrm{E}\bigg(\frac{|z(\bm{s}_i)-z(\bm{s}_j)|^{1/2}-C_1}{C_2\|\bm{s}_i-\bm{s}_j\|}\bigg)\approx\sigma_i^2/\alpha_i;\quad i=1,\dots,n,}
\label{eq:mean}
\end{equation}

\noindent \change{where $C_1\equiv 2^{1/2}\pi^{-1/2}\Gamma(3/4)(\tau^2)^{1/4}$ and $C_2\equiv 2^{-3/2}\pi^{-1/2}\Gamma(3/4)(\tau^2)^{-3/4}$.}
Note that the left-hand side of (6) depends on \change{$\|\bm{s}_j-\bm{s}_i\|$}, but the right-hand side does not.
\change{In addition, from \eqref{eq:CH}, for small $\|\bm{s}_j-\bm{s}_i\|$,}
\begin{align*}
\change{\mathrm{var}\bigg(\frac{|z(\bm{s}_i)-z(\bm{s}_j)|^{1/2}-C_1}{C_2\|\bm{s}_i-\bm{s}_j\|}\bigg)\approx
\frac{C_3}{\|\bm{s}_j-\bm{s}_i\|^2}},
\end{align*}

\noindent \change{where $C_3\equiv 2\big(\pi^{-1/2}-\pi^{-1}\Gamma(3/4)^2\big)/C_2$.}
This motivates us to use the following \change{weighted average as our local spatial indices} to monitor the heterogeneity of spatial dependence:
\begin{equation}
\change{\xi_i\equiv\frac{1}{|N_i|\sum_{j\in N_i}\omega_{ij}}\sum_{j\in N_i}\left\{\omega_{ij}
\frac{|z(\bm{s}_j)-z(\bm{s}_i)|^{1/2}-C_1}{\|\bm{s}_j-\bm{s}_i\|}\right\}};\quad i\in\mathcal{I},
\label{eq:xi}
\end{equation}

\noindent where \change{$\omega_{ij}\equiv\|\bm{s}_j-\bm{s}_i\|^2$,} $\mathcal{I}\equiv\{i:|N_i|>0,\,i=1,\dots,n\}$,
and $|N_i|$ denotes the number of elements in $N_i$\:.
\change{In practice, we recommend choosing $r=\{5|D|/(n\pi)\}^{1/2}$ in \eqref{eq:neighborhood}, so that $|N_i|\approx 5$ on average, for $i\in\mathcal{I}$.
Figures~\ref{fig:motivation}(c1) and \ref{fig:motivation}(c2) show the proposed local spatial indices of \eqref{eq:xi} based on the data in Figures~\ref{fig:motivation}(b1) and \ref{fig:motivation}(b2).
respectively.}

\change{When $\tau^{2}$ is unknown, we estimate it based on a linear extrapolation to the zero ordinate of $\gamma_{z,i}(\cdot)$ at two small lags,
determined by $\mathcal{P}_k\equiv\{(i,j):d^*_{k-1}<|\bm{s}_{i}-\bm{s}_{j}|\leq d^*_k,\,i<j\}$; $k=1,2$, where
$0=d^*_0<d^*_1<d^*_2$. 
Specifically, we compute the robust semivariogram estimates of Cressie and Hawkins (1980) based on pairs in $\mathcal{P}_k$:}
\begin{equation}
\change{\hat{\gamma}_k=\frac{\bigg(\displaystyle\sum_{(i,j)\in\mathcal{P}_k}|z(\bm{s}_i)-z(\bm{s}_j)|^{1/2}\big/m_k\bigg)^4}{2(0.457+0.494/m_k+0.045/m_k^2)};\quad k=1,2,}
\label{eq:robust}
\end{equation}

\noindent \change{where $m_k$ is the number of pairs on $\mathcal{P}_k$; $k=1,2$.
Applying linear extrapolation while imposing constraints for a nonnegative slope and intercept, we obtain}
\begin{equation}
\change{\hat{\tau}^{2}=\max\left(0,\hat{\gamma}_{1}-d_{1}\max\left(0,\frac{\hat{\gamma}_{2}-\hat{\gamma}_{1}}{d_{2}-d_{1}}\right)\right),}
\label{eq:nugget}
\end{equation}

\noindent \change{where $d_k=\sum_{(i,j)\in\mathcal{P}_k}|\bm{s}_{i}-\bm{s}_{j}|/m_k$ is the average distance among pairs in $\mathcal{P}_k$, for $k=1,2$.
Two distinct subregions are discernible from Figure~\ref{fig:fused}. In contrast, there is no clear pattern from Figure~\ref{fig:fused2}.}

\subsection{\change{Multiresolution spatial visualization}}
\label{sec:visualization}

We note from \eqref{eq:CH} that $\xi_i$ is approximately Gaussian with \change{$\mathrm{E}(\xi_i)\approx C_2\sigma^2_i/\alpha_i$}\:,
for $i\in\mathcal{I}$.
Let $n^*\equiv|\mathcal{I}|$, and without loss of generality,
assume that $\mathcal{I}=\{1,\dots,n^*\}$.
\change{If $y(\cdot)$ is globally stationary, we have $\sigma^2_1=\cdots=\sigma^2_{n^*}$ and $\alpha_1=\cdots=\alpha_{n^*}$},
and hence $\mathrm{E}(\xi_1)\approx\cdots\approx\mathrm{E}(\xi_{n^*})$.
We can apply a spatial-clustering approach to segment $D$ into stationary components based on $\bm{\xi}\equiv(\xi_1,\dots,\xi_{n^*})'$.

\begin{figure}\centering
\begin{tabular}{ccc}
\includegraphics[scale=0.75,trim={0cm 1cm 0cm 1cm},clip]{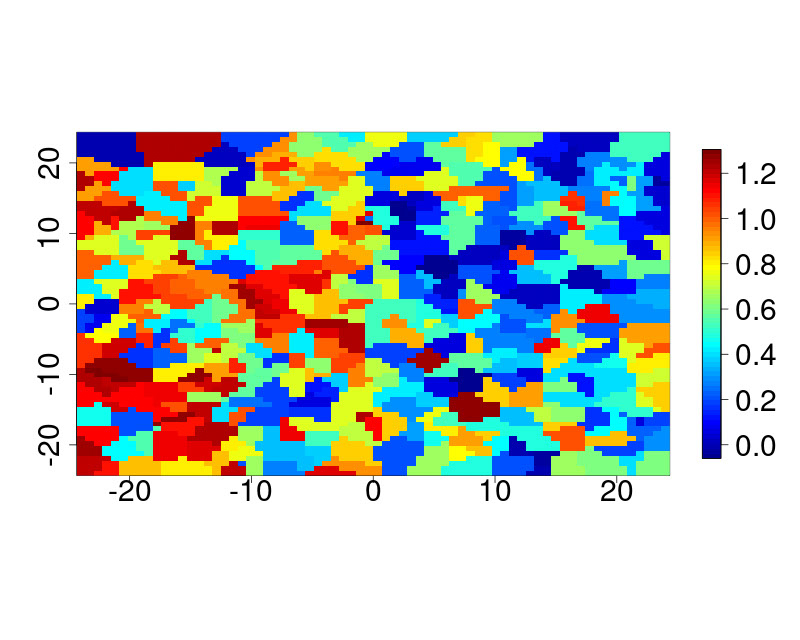} &
\includegraphics[scale=0.75,trim={0cm 1cm 0cm 1cm},clip]{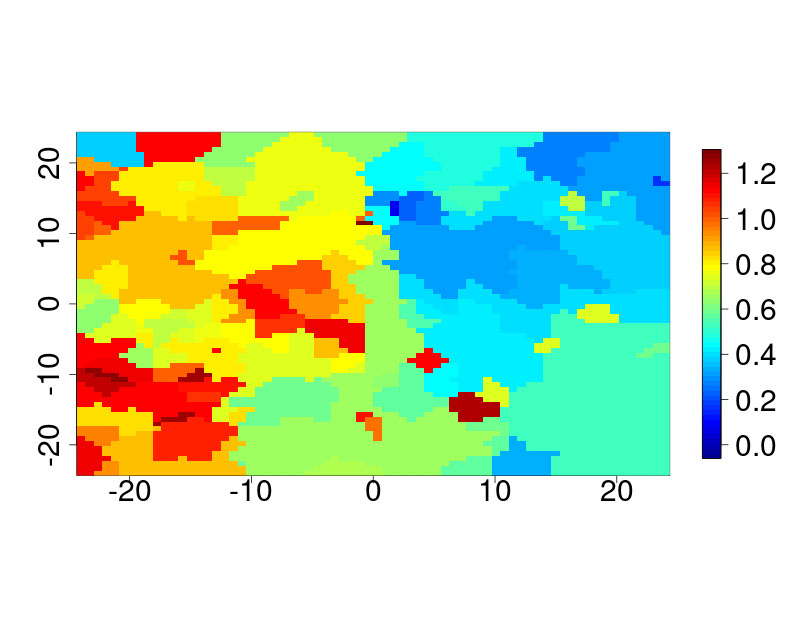} &
\includegraphics[scale=0.75,trim={0cm 1cm 0cm 1cm},clip]{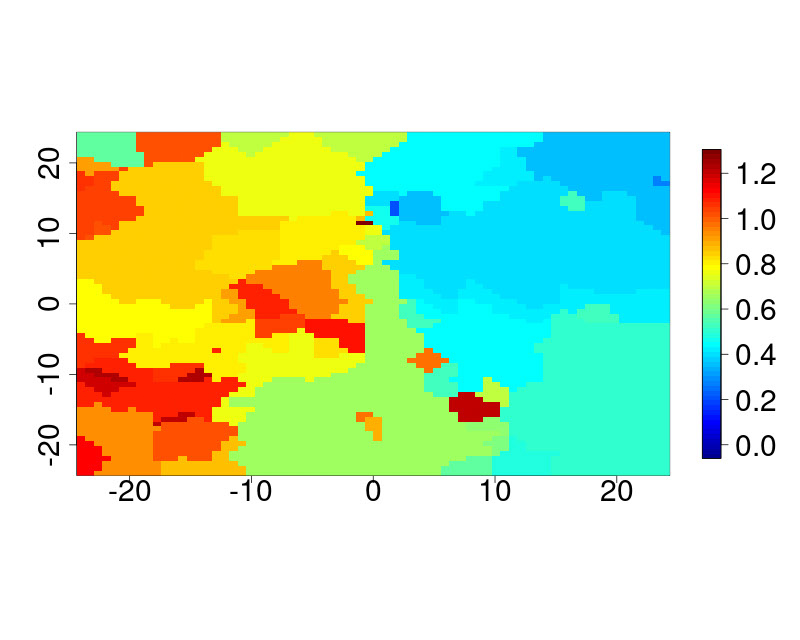} \\
\includegraphics[scale=0.75,trim={0cm 1cm 0cm 1cm},clip]{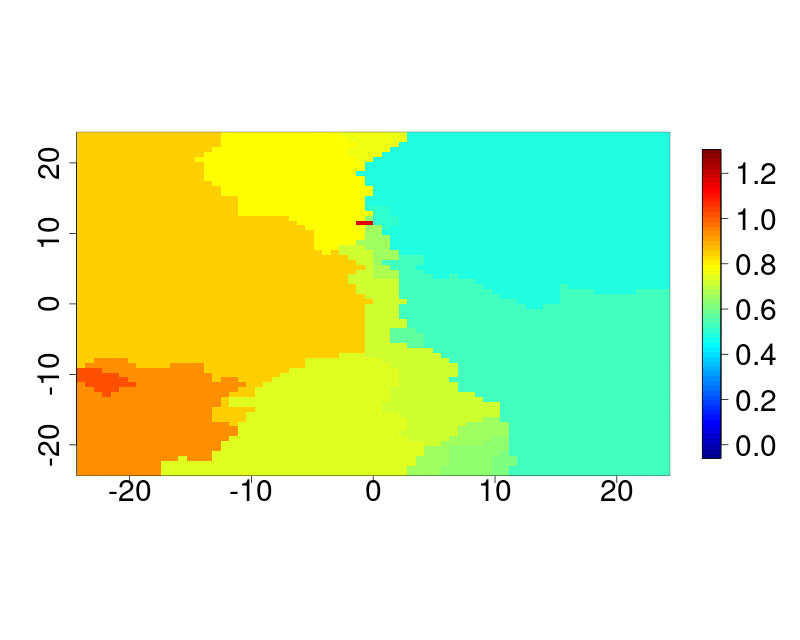} &
\includegraphics[scale=0.75,trim={0cm 1cm 0cm 1cm},clip]{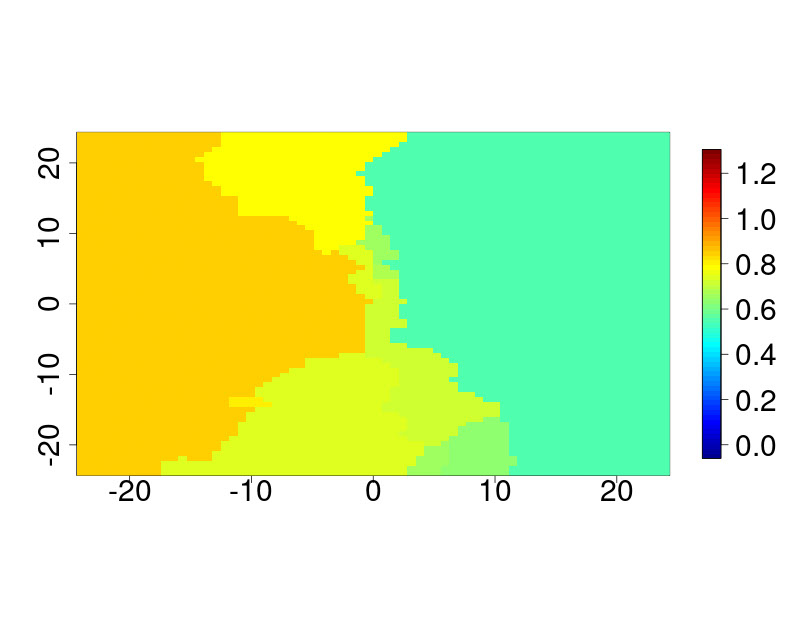} &
\includegraphics[scale=0.75,trim={0cm 1cm 0cm 1cm},clip]{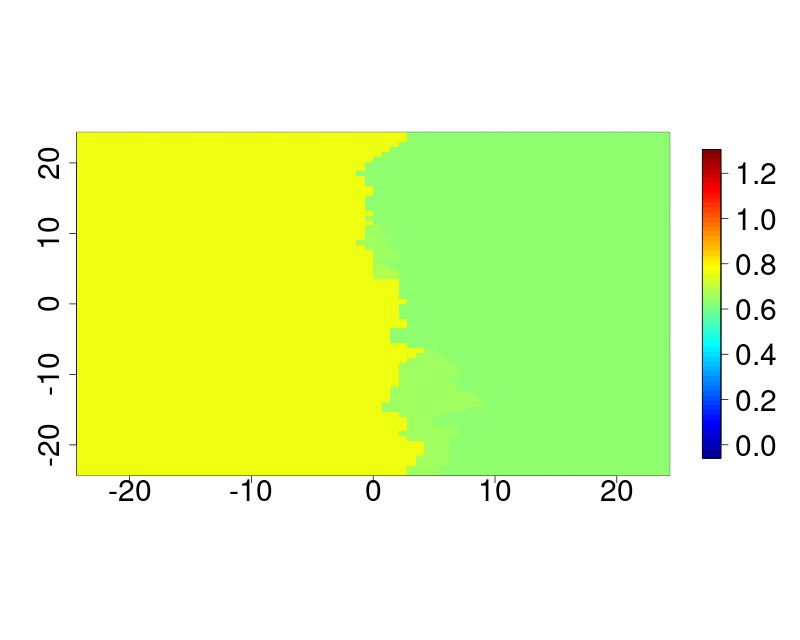}
\end{tabular}
\caption{\change{Multiresolution spatial visualization for a piecewise stationary process using the proposed fused lasso method.}}
\label{fig:fused}
\end{figure}

\begin{figure}\centering
\begin{tabular}{ccc}
\includegraphics[scale=0.75,trim={0cm 1cm 0cm 1cm},clip]{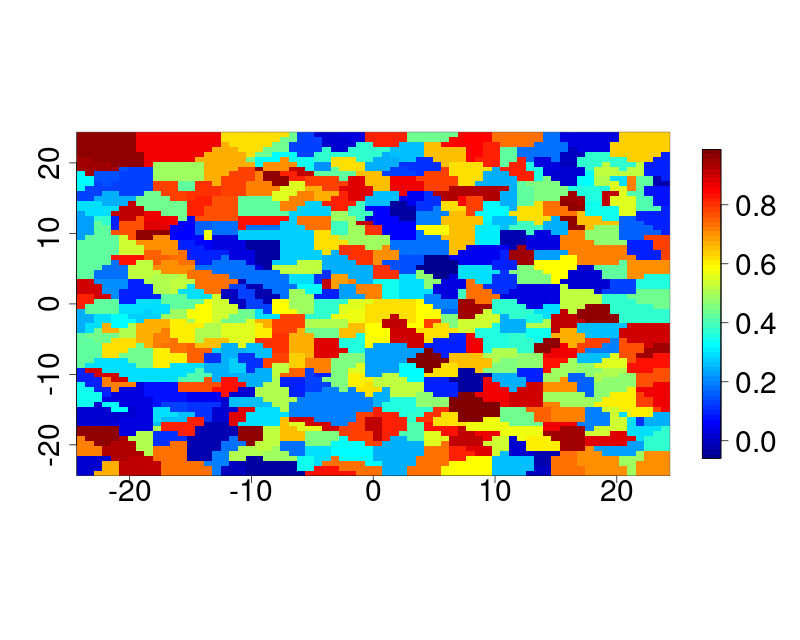} &
\includegraphics[scale=0.75,trim={0cm 1cm 0cm 1cm},clip]{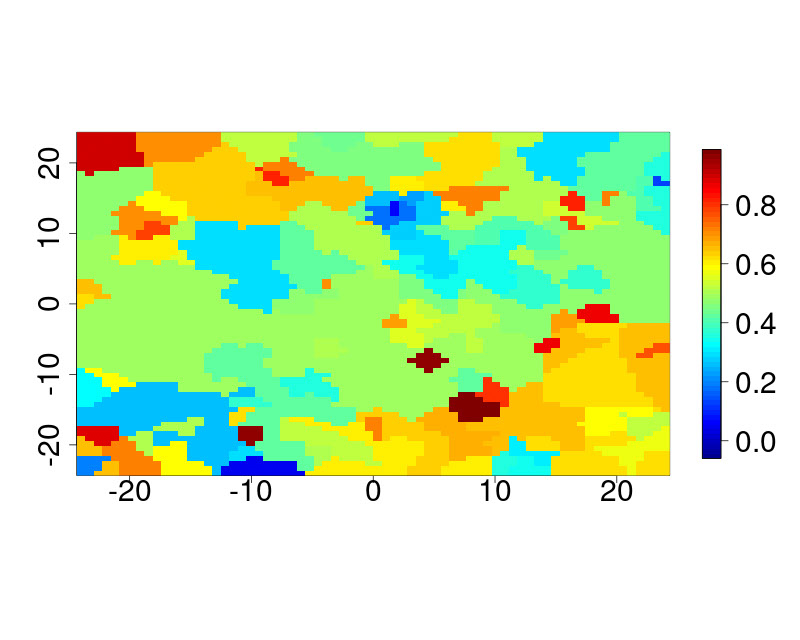} &
\includegraphics[scale=0.75,trim={0cm 1cm 0cm 1cm},clip]{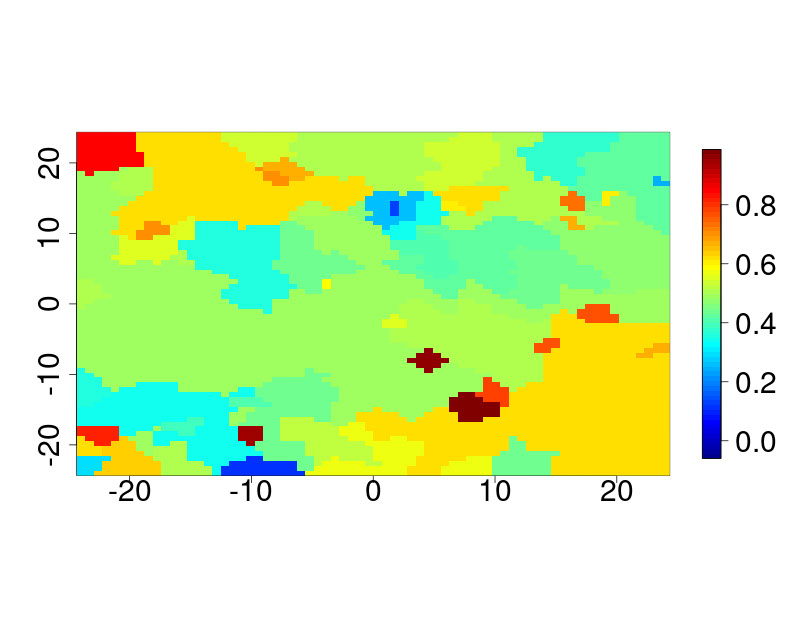} \\
\includegraphics[scale=0.75,trim={0cm 1cm 0cm 1cm},clip]{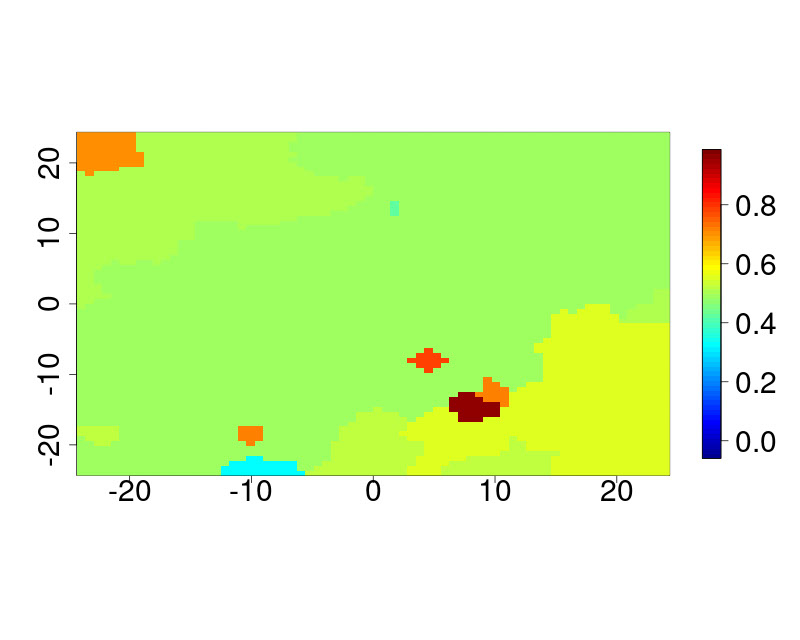} &
\includegraphics[scale=0.75,trim={0cm 1cm 0cm 1cm},clip]{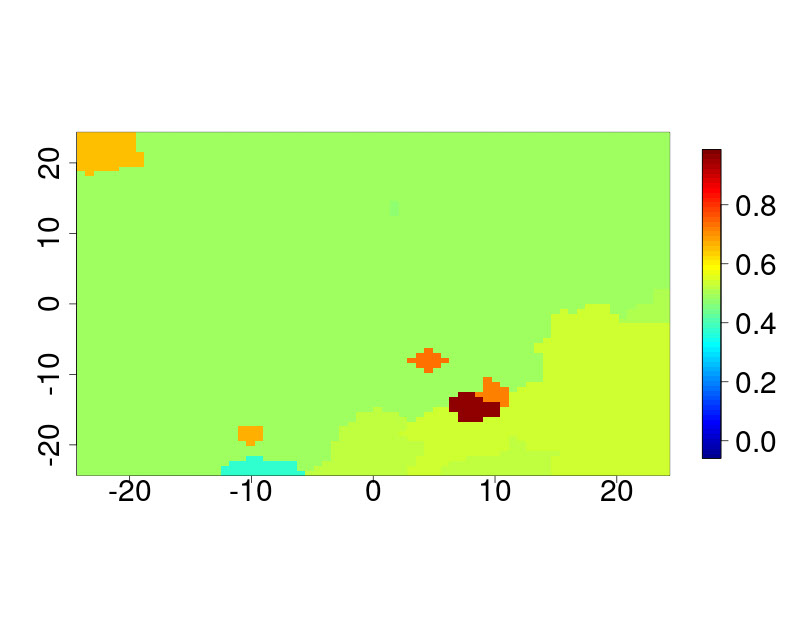} &
\includegraphics[scale=0.75,trim={0cm 1cm 0cm 1cm},clip]{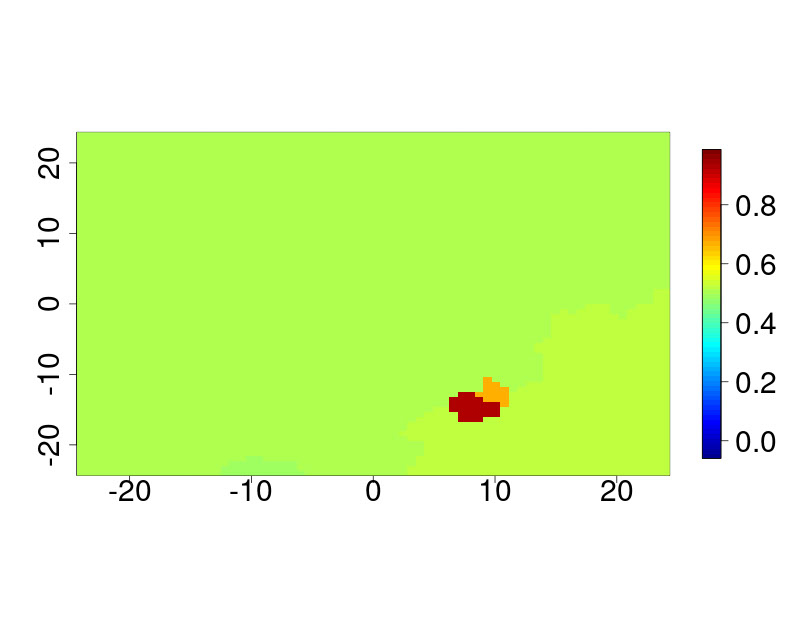}
\end{tabular}
\caption{\change{Multiresolution spatial visualization for a stationary process using the proposed fused lasso method.}}
\label{fig:fused2}
\end{figure}

\change{To explore the spatial nonstationarity evident in the data presented in Figure~\ref{fig:motivation}(b2) at various resolutions,
we decompose $D$ into disjoint Voronoi cells, denoted as $\{B_1,\dots,B_{n^*}\}$, corresponding to $\{\bm{s}_1,\dots,\bm{s}_{n^*}\}$.
Each point in the Voronoi cell $B_k$ is closer to $\bm{s}_k$ than any other point in the set $\{\bm{s}_1,\dots,\bm{s}_{n^*}\}$.
Then we cluster $\{B_1,\dots,B_{n^*}\}$ into homogeneous components using the fused lasso (Tibshirani \textit{et al}., 2005):}
\[
\change{\sum_{i=1}^{n^*}(\xi_i-\beta_i)^2+\rho\sum_{(j,k)\in\mathcal{E}}|\beta_j-\beta_k|.}
\]
\change{where $\mathcal{E}$ is obtained by linking between any two cells that share a boundary and $\rho\geq 0$ is a regularization parameter.
The resulting images at multiple resolutions with different tuning parameter values of $\rho$ are shown in Figure~\ref{fig:fused}.
The corresponding images for the data in Figure~\ref{fig:motivation}(b1) from a stationary process are shown in Figure~\ref{fig:fused2}.
Comparing the two collections of images,
it is evident that Figure~\ref{fig:fused} showcases at least two major components with distinct values across different resolutions. In contrast, Figure~\ref{fig:fused2} consistently exhibits a single prominent homogeneous component throughout all resolutions.}

\subsection{Voronoi tessellations into stationary components}
\label{sec:Voronoi}

\change{The fused lasso approach given in the previous section is used for visualization.
We aim to partition the region $D$ into stationary components for subsequent spatial prediction.
To achieve this, we utilize the Voronoi subregions constructed from $K$ seeds denoted as $\{\bm{p}_1,\dots,\bm{p}_K\}\subset\mathbb{R}^2$.
Given these seeds, we derive the corresponding Voronoi tessellation, subdividing $D$ into $K$ distinct components $D_1,\dots,D_K$.
Define $n_k\equiv\sum_{i\in\mathcal{I}} I(\bm{s}_i\in D_k)$; $k=1,\dots,K$.
Let $\mathcal{S}_K$ be the set of all possible $K$ seeds.
To identify the optimal set of $K$ seeds $\{\bm{p}_1,\dots,\bm{p}_K\}$ from $\mathcal{S}_k$,
we apply the following objective function grounded in independent normal likelihood:}
\begin{equation}
f_K(\bm{p}_1,\dots,\bm{p}_K;\bm{\xi})=\change{\sum_{k=1}^K\sum_{i\in\mathcal{I}:\bm{s}_i\in D_k}}
\bigg\{\log(\hat{v}_k(D_k))-\log\phi\bigg(\frac{\xi_i-\bar{\mu}_k(D_k)}{\hat{v}_k(D_k)}\bigg)\bigg\},
\label{eq:obj}
\end{equation}

\noindent where
\[
\bar{\mu}_k(D_k)\equiv\frac{1}{n_k}\sum_{i\in\mathcal{I}:\bm{s}_i\in D_k}\xi_i,\quad\mbox{and}\quad
\hat{v}^2_k(D_k)\equiv\frac{1}{n_k}\sum_{i\in\mathcal{I}:\bm{s}_i\in D_k}(\xi_i-\bar{\mu}_k(D_k))^2,
\] 
are the maximum likelihood (ML) estimators of the mean and the variance of \change{$\{\xi_i:\bm{s}_i\in D_k,\,i\in\mathcal{I}\}$, for $k=1,\dots,K$},
and $\phi(\cdot)$ is the probability density function of the standard normal distribution.
\change{The proposed segmentation of $D$ into $K\geq 2$ components is determined by}
\begin{equation}
\big\{\hat{\bm{p}}_1^{(K)},\dots,\hat{\bm{p}}_K^{(K)}\big\}\equiv\mathop{\arg\,\min}_{\{\bm{p}_1,\dots,\bm{p}_K\}\in\mathcal{S}_K}f_K(\bm{p}_1,\dots,\bm{p}_K;\bm{\xi})
\label{eq:partition}
\end{equation}

\noindent with the corresponding Voronoi tessellation $\hat{D}_1^{(K)},\dots,\hat{D}_K^{(K)}$. 

\change{We propose a simple algorithm to find the solution of \eqref{eq:partition}.}
Its pseudo-code is outlined in Algorithm~\ref{alg:f}.
\begin{algorithm}[h]
\caption{\change{Find the solution of \eqref{eq:partition} with a given $K$ based on $\bm{\xi}$.}}
\label{alg:f}
  \begin{algorithmic}
    \Require
    \State \change{$\{\bm{p}_1,\dots,\bm{p}_K\}$: the initial seeds obtained from a deterministic $K$-means algorithm of Nidheesh \textit{et al}.~(2017);}
    \State \change{$\{D_1,\dots,D_K\}$: the Voronoi tessellations corresponding to $\{\bm{p}_1,\dots,\bm{p}_K\}$.}
    \Ensure
    \Repeat
      \For{$k \gets 1$ to $K$}             
        \State \multiline{%
          update $\bm{p}_k$ by replacing it from $\{\bm{s}_i\in D_k:i=1,\dots,n^*\}$ such that
          $f_K(\bm{p}_1,\dots,\bm{p}_K;\bm{\xi})$ is minimized;}
        \State update $\{D_1,\dots,D_K\}$ corresponding to the current seeds $\{\bm{p}_1,\dots,\bm{p}_K\}$;
      \EndFor
    \Until{no further reduction of $f_K(\bm{p}_1,\dots,\bm{p}_K;\bm{\xi})$ is possible.}
  \end{algorithmic}
\end{algorithm}

\section{The proposed test for stationarity}
\label{sect:test}

\change{We can utilize the proposed segmentation with $K=2$ to test for spatial stationarity.}
We consider the following hypothesis test:
\[
H_0: y(\cdot)\mbox{ is stationary~~versus~~}H_1: y(\cdot)\mbox{ is not stationary}.
\]
Based on the two subregions $\hat{D}_1^{(2)}$ and $\hat{D}_2^{(2)}$ selected by \eqref{eq:partition} with $K=2$,
\change{we propose the following two-sample $t$ statistic}:
\begin{equation}
T=\frac{\big|\bar{\mu}_1(\hat{D}_1^{(2)})-\bar{\mu}_2(\hat{D}_2^{(2)})\big|}
  {\sqrt{\hat{v}^2_1(\hat{D}_1^{(2)})/(\change{\hat{n}_1}-1)+\hat{v}^2_2(\hat{D}_2^{(2)})/(\hat{n}_2-1)}},
\label{eq:test}
\end{equation}

\noindent where $\hat{n}_k\equiv\sum_{i=1}^{n^*} I(\bm{s}_i\in \hat{D}_k^{(2)})$; $k=1,2$.
The distribution of $T$ is complicated because there is a selection process involved in obtaining $\hat{D}_1^{(2)}$ and $\hat{D}_2^{(2)}$.
So, we apply a Monte Carlo (MC) method to find \change{the null distribution of $T$}.
Specifically, we assume that under $H_0$,
$y(\cdot)$ is a Gaussian process with the isotropic Mat\'{e}rn covariance model of \eqref{eq:Matern}.
We estimate $\bm{\theta}$ and $\change{\sigma^2}$ in \eqref{eq:Matern} by ML.
\change{The ML estimator of  $\bm{\theta}=(\alpha,\nu)'$ can be obtained} by minimizing the negative log profile likelihood:
\begin{equation}
\hat{\bm{\theta}}\,\equiv\,\big(\hat{\alpha},\hat{\nu}\big)'\,\equiv\,
\mathop{\arg\min}_{\bm{\theta}}\bigg\{\frac{1}{2}\log|\bm{\Omega}(\bm{\theta})|
+\frac{K}{2}\log\big\{\bm{z}'\bm{\Omega}(\bm{\theta})^{-1}\bm{z}\big\}+\mathrm{constant}\bigg\},
\label{eq:gamma}
\end{equation}

\noindent where $\bm{\Omega}(\bm{\theta})$ is an $n\times n$ correlation matrix whose $(i,j)$-th entry is
\change{$\{\mathrm{cov}(y(\bm{s}_i),y(\bm{s}_j))+\tau^2\delta_{ij}\}/\sigma^2$}.
Then the ML estimator of $\change{\sigma^2}$ is given by:
\begin{equation}
\change{\hat{\sigma}^2}\,\equiv\,\frac{1}{n}\bm{z}'\bm{\Omega}(\hat{\bm{\theta}})^{-1}\bm{z}.
\label{eq:tau}
\end{equation}

\noindent To implement the proposed MC method, first, we simulate data $\bm{z}^{(m)}$, for $m=1,\dots,M$, based on \eqref{eq:data} and \eqref{eq:Matern}
with $\bm{\theta}$ and $\change{\sigma^2}$ replaced by $\hat{\bm{\theta}}$ in \eqref{eq:gamma} and $\change{\hat{\sigma}^2}$ in \eqref{eq:tau}.
Next, we compute $T_m$ in \eqref{eq:test} based on $\bm{z}^{(m)}$.
Then the MC $p$-value of the proposed test is
\begin{equation}
\hat{p}=\frac{1}{M+1}\sum_{m=1}^M I(T_m>T).
\label{eq:pvalue}
\end{equation}

\change{Although we introduce the segmentation method before hypothesis testing, in practice,}
we first perform the stationarity test and obtain the $p$-value $\hat{p}$ of \eqref{eq:pvalue}.
We use a stationary model for subsequent analysis if $\hat{p}\geq 0.05$.
Otherwise, we apply the proposed spatial segmentation method to partition $D$ into $K$ stationary subregions
with $K$ selected by \change{minimizing} Bayesian information criterion (BIC) (Schwarz, 1978):
\begin{equation}
\mathrm{BIC}(K)=f_K\big(\hat{\bm{p}}_1^{(K)},\dots,\hat{\bm{p}}_K^{(K)}\big)+4K\log(n^*).
\label{eq:BIC}
\end{equation}


\section{Simulation studies}

\subsection{Testing stationarity}
\label{sec:test}

We examined the size of the proposed stationarity test under $H_0$ by performing the same simulation experiment as
in Section 7.1.1 of Bandyopadhyay and Rao (2017). 
We considered a zero-mean spatial process $\{y(\bm{s}):\bm{s}\in D\}$ on a region $D=[-5/2,5/2]\times[-5/2,5/2]$
with a Mat\'{e}rn covariance function of \eqref{eq:Matern}.
We generated data according to \eqref{eq:data} with $\change{\sigma^2}=1$, $\nu=1/2$, $\alpha\in\{1/3,2/3,1,4/3,2\}$,
and $\change{\tau^2}\in\{0,0.01\}$.
In addition, we considered various sample sizes $n\in\{50,100,500,1000,2000\}$ and two distributions for sampling locations,
including a uniform distribution and a clustered distribution with two clusters (see details in Bandyopadhyay and Rao, 2017),
resulting in a total of $5 \times 2 \times 5 \times 2 $=100 combinations.

We compared our method with BR's (Bandyopadhyay and Rao, 2017).
The empirical Type-I error rates under various settings for the uniform and the clustered distributions are shown in
Table \ref{alpha-uniform} and Table \ref{alpha-clustered}, respectively.
Although our method shows a few elevated Type-I error rates when spatial dependence is strong,
overall, the Type-I error rates are close to the nominal level.
On the other hand, the Type-I error rates for the BR's method tend to be too large for a few cases
under the uniform design and too small for many instances under the clustered design.
The distributions of p-values for various scenarios \change{under $H_0$} are displayed in Figures \ref{fig:pval-1a}-\ref{fig:pval-2b}.
They are all very close to the uniform distribution on $(0,1)$ as we anticipate.

\begin{table}\centering
\caption{Empirical Type-I errors for our method and BR's method (Bandyopadhyay and
Rao, 2017) under various scenarios with a uniform sampling design based on 500 simulated replicates.}
\begin{small}
\begin{tabular}{cc|lllll|rrrrr}
\hline 
$n$ & Method & \multicolumn{5}{c|}{$\change{\tau^2}=0$} & \multicolumn{5}{c}{$\change{\tau^2}=0.01$}\tabularnewline
\cline{3-12}
& & $\alpha=\frac{1}{3}$\! & $\alpha=\frac{2}{3}$\! & $\alpha=1$ & $\alpha=\frac{4}{3}$\! & $\alpha=2$ &
  $\alpha=\frac{1}{3}$\! & $\alpha=\frac{2}{3}$\! & $\alpha=1$ & $\alpha=\frac{4}{3}$\! & $\alpha=2$\tabularnewline
\cline{1-12}
50   & Ours & 0.068 & 0.054 & 0.052 & 0.058 & 0.060 & 0.062 & 0.060 & 0.056 & 0.058 & 0.054\tabularnewline
50   & BR    & 0.030 & 0.020 & 0.040 & 0.050 & 0.090 & 0.030 & 0.020 & 0.050 & 0.050 & 0.080\tabularnewline
100  & Ours & 0.034 & 0.042 & 0.044 & 0.054 & 0.058 & 0.034 & 0.046 & 0.044 & 0.056 & 0.050\tabularnewline
100  & BR    & 0.030 & 0.030 & 0.030 & 0.040 & 0.040 & 0.030 & 0.050 & 0.040 & 0.050 & 0.040\tabularnewline
500  & Ours & 0.046 & 0.054 & 0.056 & 0.052 & 0.050 & 0.040 & 0.050 & 0.048 & 0.054 & 0.048\tabularnewline
500  & BR    & 0.020 & 0.030 & 0.020 & 0.030 & 0.030 & 0.040 & 0.080 & 0.070 & 0.130 & 0.120\tabularnewline
1000 & Ours & 0.072 & 0.072 & 0.068 & 0.070 & 0.054 & 0.072 & 0.072 & 0.070 & 0.070 & 0.068\tabularnewline
1000 & BR    & 0.050 & 0.050 & 0.060 & 0.060 & 0.080 & 0.080 & 0.100 & 0.100 & 0.140 & 0.130\tabularnewline
2000 & Ours & 0.078 & 0.066 & 0.064 & 0.064 & 0.072 & 0.074 & 0.074 & 0.068 & 0.076 & 0.080\tabularnewline
2000 & BR    & 0.070 & 0.060 & 0.090 & 0.090 & 0.080 & 0.070 & 0.090 & 0.090 & 0.180 & 0.180\tabularnewline
\hline 
\end{tabular}
\end{small}
\label{alpha-uniform}
\end{table}

\begin{table}\centering
\caption{Empirical Type-I errors for our method and BR's method (Bandyopadhyay and
Rao, 2017) under various scenarios with a clustered sampling design based on 500 simulated replicates.}
\begin{small}
\begin{tabular}{cc|rrrrr|rrrrr}
\hline 
$n$ & Method & \multicolumn{5}{c|}{$\change{\tau^2}=0$} & \multicolumn{5}{c}{$\change{\tau^2}=0.01$}\tabularnewline
\cline{3-12}
& & $\alpha=\frac{1}{3}$\! & $\alpha=\frac{2}{3}$\! & $\alpha=1$ & $\alpha=\frac{4}{3}$\! & $\alpha=2$ &
  $\alpha=\frac{1}{3}$\! & $\alpha=\frac{2}{3}$\! & $\alpha=1$ & $\alpha=\frac{4}{3}$\! & $\alpha=2$\tabularnewline
\cline{1-12}
  50 & ours & 0.064 & 0.060 & 0.074 & 0.072 & 0.078 & 0.058 & 0.06 & 0.070 & 0.076 & 0.072\tabularnewline
  50 & BR & 0.020 & 0.020 & 0.010 & 0.010 & 0.020 & 0.020 & 0.020 & 0.020 & 0.030 & 0.020\tabularnewline
 100 & ours & 0.082 & 0.074 & 0.072 & 0.062 & 0.054 & 0.080 & 0.078 & 0.064 & 0.074 & 0.068\tabularnewline
 100 & BR & 0.020 & 0.020 & 0.020 & 0.020 & 0.020 & 0.020 & 0.030 & 0.020 & 0.030 & 0.030\tabularnewline
500 & ours & 0.060 & 0.052 & 0.058 & 0.060 & 0.056 & 0.064 & 0.052 & 0.052 & 0.060 & 0.060\tabularnewline
500 & BR & 0.020 & 0.020 & 0.010 & 0.010 & 0.010 & 0.010 & 0.020 & 0.020 & 0.010 & 0.010\tabularnewline
1000 & ours & 0.056 & 0.058 & 0.064 & 0.066 & 0.060 & 0.060 & 0.060 & 0.060 & 0.062 & 0.062\tabularnewline
1000 & BR & 0.010 & 0.010 & 0.010 & 0.010 & 0.010 & 0.010 & 0.010 & 0.010 & 0.010 & 0.020\tabularnewline
2000 & ours & 0.052 & 0.062 & 0.074 & 0.076 & 0.076 & 0.058 & 0.058 & 0.072 & 0.074 & 0.076\tabularnewline
2000 & BR & 0.010 & 0.010 & 0.010 & 0.010 & 0.010 & 0.010 & 0.010 & 0.010 & 0.010 & 0.010\tabularnewline
\hline 
\end{tabular}
\par\end{small}
\label{alpha-clustered}
\end{table}

Next, we investigated the power of the proposed test following the same setups in Bandyopadhyay and Rao (2017).
We considered three scenarios. In the first two scenarios,
we replaced the stationary Mat\'{e}rn covariance function of \eqref{eq:Matern} by a nonstationary Mat\'ern covariance function
with $\lambda=20$ and $40$, respectively:
\begin{equation}
\mathrm{cov}(y(\bm{s}_1),y(\bm{s}_2))=|\bm{\Sigma}_\lambda(\bm{s}_1)|^{1/4}|\bm{\Sigma}_\lambda(\bm{s}_2)|^{1/4}
|\{\bm{\Sigma}_\lambda(\bm{s}_1)+\bm{\Sigma}_\lambda(\bm{s}_2)\}/2|^{-1/2}
\exp\big(-\sqrt{Q_\lambda(\bm{s}_1,\bm{s}_2)}\big),
\label{eq:smoothly-nonstationary}
\end{equation}

\noindent where
\begin{align*}
  Q_{\lambda}(\bm{s}_{1},\bm{s}_{2})
\equiv&~ 2(\bm{s}_{1}-\bm{s}_{2})'\big\{\bm{\Sigma}_\lambda(\bm{s}_1)+\bm{\Sigma}_\lambda(\bm{s}_2)\big\}^{-1}(\bm{s}_{1}-\bm{s}_{2}),\\
  \bm{\Sigma}_\lambda(\bm{s})
\equiv&~ \left(
  \begin{matrix}
  \log\Big(\displaystyle\frac{s_x}{\lambda}+\frac{3}{4}\Big) & -\displaystyle\frac{\|\bm{s}\|^2}{\lambda^2}\\
 \displaystyle\frac{\|\bm{s}\|^2}{\lambda^2} & \log\Big(\displaystyle\frac{s_x}{\lambda}+\frac{3}{4}\Big)
  \end{matrix}
  \right)\left(
  \begin{matrix}
  1 & 0\\
  0 & 0.5
  \end{matrix}
  \right)\left(
  \begin{matrix}
  \log\Big(\displaystyle\frac{s_x}{\lambda}+\frac{3}{4}\Big) & \displaystyle\frac{\|\bm{s}\|^2}{\lambda^2}\\
  -\displaystyle\frac{\|\bm{s}\|^2}{\lambda^2} & \log\Big(\displaystyle\frac{s_x}{\lambda}+\frac{3}{4}\Big)
  \end{matrix}
  \right),
\end{align*}
 
\noindent and $\bm{s}=(s_x,s_y)'$.
For the third scenario, we considered a zero-mean piecewise stationary process $\{y(\bm{s}):\bm{s}\in D\}$
by dividing $D=[-5/2,5/2]\times[-5/2,5/2]$ into $2\times 2$ blocks of equal sizes.
The processes on four blocks are mutually independent and have the Mat\'{e}rn covariance functions of (\ref{eq:Matern}),
with $\change{\sigma^2}=1$, $\nu=1/2$, and four different values of $\alpha\in\{1,1/3,1/2,2/3\}$ for the four blocks.
For each scenario, we considered the uniform sampling design and generated data according to \eqref{eq:data}
with $\change{\tau^2}\in\{0,0.01\}$ and $n\in\{50,100,500,1000,2000\}$, resulting in 10 different combinations.
The empirical powers are displayed in Table \ref{power} based on 500 simulated replicates. 
Except for a few cases in Scenario 2 with $\gamma=40$ and $n\geq 1000$,
\change{our method is more powerful than the BR's method in detecting spatial nonstationarity.}

\begin{table}[tb]\centering
\caption{Empirical powers for our and BR's methods (Bandyopadhyay and Rao, 2017) under various scenarios based on 500 simulated replicates.}
\begin{tabular}{cc|rrr|rrr}
\hline 
$n$ & Method & \multicolumn{3}{c|}{$\change{\tau^2}=0$} & \multicolumn{3}{c}{$\change{\tau^2}=0.01$}\tabularnewline
\cline{1-8}
& & $\lambda=20$ & $\lambda=40$ & 4 blocks & $\lambda=20$ & $\lambda=40$ & 4 blocks\tabularnewline
\cline{3-8} \cline{4-8} \cline{5-8} \cline{6-8} \cline{7-8} \cline{8-8} 
50 & Our & 0.050 & 0.058 & 0.074 & 0.054 & 0.060 & 0.076\tabularnewline
50 & BR & 0.020 & 0.030 & 0.050 & 0.030 & 0.040 & 0.050\tabularnewline
100 & Our & 0.060 & 0.044 & 0.106 & 0.058 & 0.048 & 0.100\tabularnewline
100 & BR & 0.050 & 0.040 & 0.040 & 0.040 & 0.040 & 0.040\tabularnewline
500 & Our & 0.340 & 0.136 & 0.570 & 0.328 & 0.140 & 0.548\tabularnewline
500 & BR & 0.190 & 0.100 & 0.110 & 0.180 & 0.090 & 0.110\tabularnewline
1000 & Our & 0.760 & 0.266 & 0.926 & 0.744 & 0.264 & 0.910\tabularnewline
1000 & BR & 0.470 & 0.350 & 0.240 & 0.460 & 0.360 & 0.240\tabularnewline
2000 & Our & 0.990 & 0.570 & 1.000 & 0.980 & 0.560 & 1.000\tabularnewline
2000 & BR & 0.700 & 0.850 & 0.360 & 0.710 & 0.850 & 0.360\tabularnewline
\hline 
\end{tabular}
\label{power}
\end{table}

\subsection{Spatial segmentation}

\begin{figure}[tb]\centering
\includegraphics[scale=0.55]{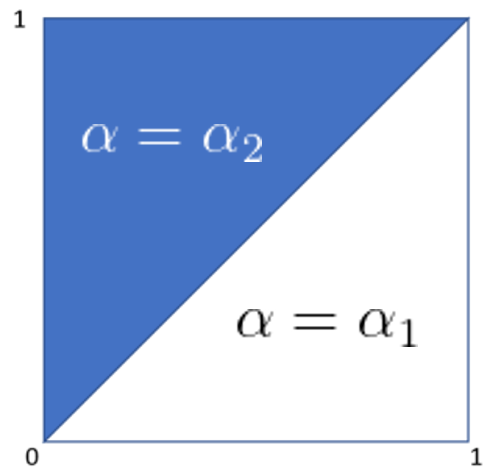}
\caption{A partition of $D$ into $D_1$ and $D_2$ and their corresponding spatial dependence parameters $\alpha_1$ and $\alpha_2$.}
\label{fig:partition2}
\end{figure}

\begin{figure}[tb]\centering
\begin{tabular}{ccccc}
$\alpha_2=0.1$ & $\alpha_2=0.2$ & $\alpha_2=0.3$ & $\alpha_2=0.4$ & $\alpha_2=0.5$\\
\includegraphics[scale=0.17,trim={2cm 1.5cm 1.5cm 0cm},clip]{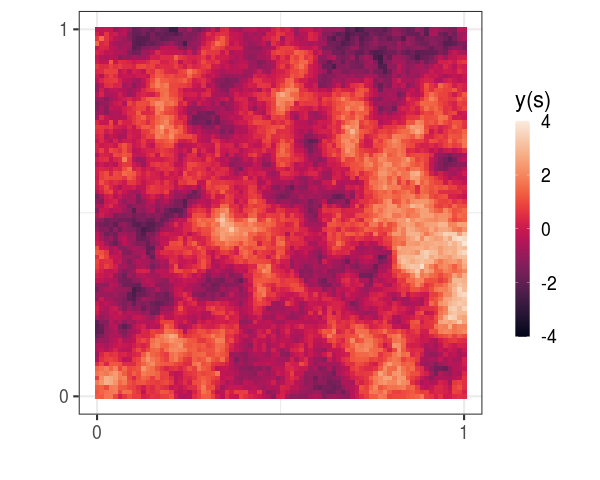} &
\includegraphics[scale=0.17,trim={2cm 1.5cm 1.5cm 0cm},clip]{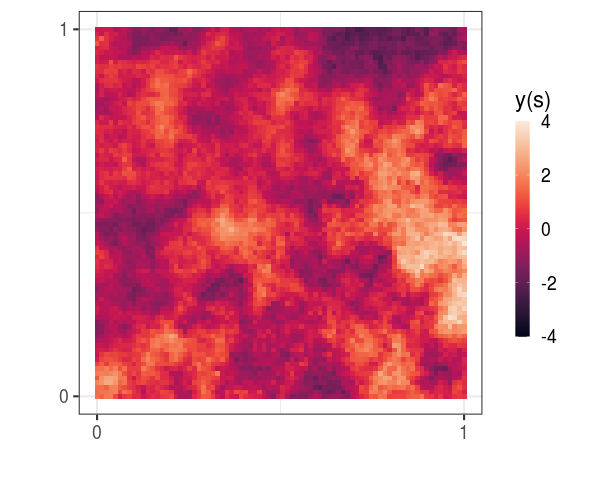} &
\includegraphics[scale=0.17,trim={2cm 1.5cm 1.5cm 0cm},clip]{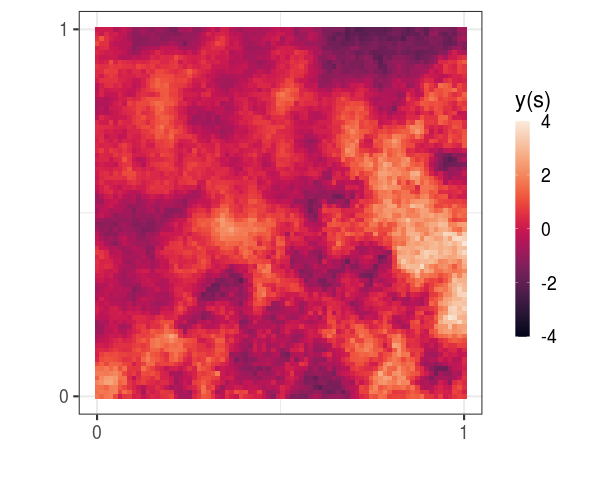} &
\includegraphics[scale=0.17,trim={2cm 1.5cm 1.5cm 0cm},clip]{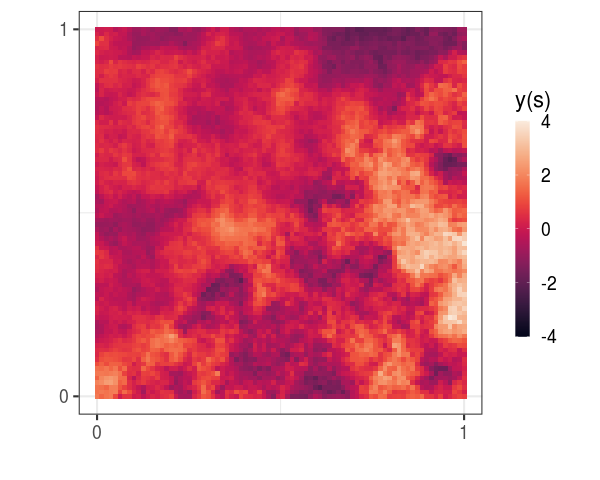} &
\includegraphics[scale=0.17,trim={2cm 1.5cm 1.5cm 0cm},clip]{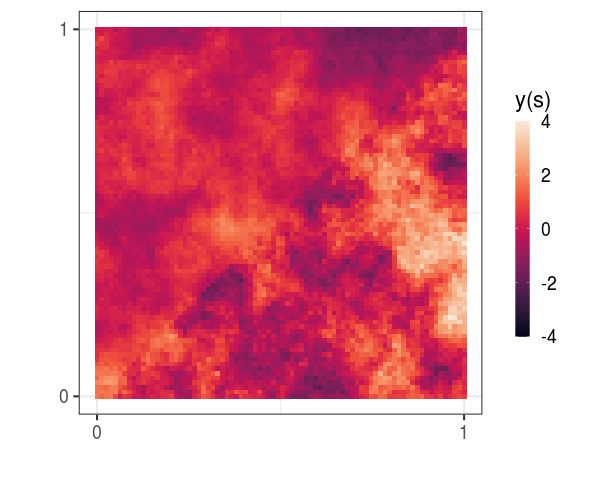}
\end{tabular}
\caption{Realizations of $y(\cdot)$ from models with various $\alpha_2$ values,
where a larger $\alpha_2$ value corresponds to a higher degree of nonstationarity,
and $\alpha_2=0.1$ corresponds to a stationary process.}
\label{fig:realization}
\end{figure}

\change{We investigated the cluster recovery ability of the proposed method in spatial segmentation.}
We considered a region $D=[0,1]^2$ and decomposed it into $D_1\cup D_2$ as shown in Figure \ref{fig:partition2}.
We generated a zero-mean spatial process $\{y(\bm{s}):\bm{s}\in D\}$ on $D$ based on
\begin{equation}
y(\bm{s})=w_1(\bm{s};a)\eta_1(\bm{s})+w_2(\bm{s};a)\eta_2(\bm{s});\quad\bm{s}\in D,
\label{eq:y2}
\end{equation}

\noindent where
\[
w_k(\bm{s};a)\equiv\frac{\exp(-d(\bm{s},D_k)/a)}{\exp(-d(\bm{s},D_1)/a)+\exp(-d(\bm{s},D_2)/a)};\quad\bm{s}\in D,\,k\in\{1,2\},
\]
are weight functions with $a>0$ controlling the degree of smoothness for process $y(\cdot)$ around the boundary between $D_1$ and $D_2$,
$d(\bm{s},D_k)\equiv\displaystyle\min_{\bm{s}^*\in D_k}\|\bm{s}-\bm{s}^*\|$,
and $\bm{\eta}(\bm{s})\equiv(\eta_1(\bm{s}),\eta_2(\bm{s}))'$ is a zero-mean bivariate spatial process with a bivariate exponential covariance function:
\[
\mathrm{cov}(\eta_k(\bm{s}),\eta_{k'}(\bm{s}^*))=\bigg(\frac{2\alpha_k\alpha_{k'}}{\alpha_k^2+\alpha_{k'}^2}\bigg)^{1/2}
\exp\bigg(-\frac{(\alpha_k^2+\alpha_{k'}^2)^{1/2}}{2^{1/2}\alpha_k\alpha_{k'}}\|\bm{s}-\bm{s}^*\|\bigg);\quad k,k'\in\{1,2\}.
\]
We generated data according to \eqref{eq:data} and \eqref{eq:y2} with $\change{\tau^2}=0$, $\alpha_1=0.1$,
and $\alpha_2\in\{0.1,0.2,0.3,0.4,0.5\}$.
Additionally, we considered $n\in\{100,500\}$ and $a\in\{0.01,0.1\}$, resulting in a total of 20 combinations.
Note that $\alpha_2$ controls the degree of nonstationarity.
When $\alpha_1=\alpha_2$, we obtain $y(\cdot)$ to be a stationary process with
$\mathrm{cov}(y(\bm{s}),y(\bm{s}^*))=\exp(-\|\bm{s}-\bm{s}^*\|/\alpha_1)$ regardless of the value of $a$.
By contrast, a larger departure of $\alpha_2$ from $\alpha_1$ indicates a higher degree of nonstationarity.
These features can be seen in Figure \ref{fig:realization}, which shows realizations of $y(\cdot)$ with $\alpha_2\in\{0.1,0.2,0.3,0.4,0.5\}$.

We applied the proposed \change{optimization} of \eqref{eq:partition} to segment $D$ into Voronoi subregions $\big\{\hat{D}_1,\dots,\hat{D}_K\big\}$.
We selected the final $K$ according to BIC of \eqref{eq:BIC}.
The performance of an estimated clustering $\tilde{\mathcal{D}}=\big\{\tilde{D}_1,\dots,\tilde{D}_K\big\}$ is evaluated using the Rand index (Rand, 1971)
based on $\{\bm{s}_1,\dots,\bm{s}_n\}$:
\[
R\big(\mathcal{D},\tilde{\mathcal{D}}\big)\equiv\frac{n_{00}+n_{11}}{n_{00}+n_{01}+n_{10}+n_{11}},
\]
where $\mathcal{D}=\{D_1,D_2\}$ is the true clustering,
\begin{itemize}
\item[$n_{00}$] is the number of point pairs that are in different clusters under both $\mathcal{D}$ and $\tilde{\mathcal{D}}$,
\item[$n_{01}$] is the number of point pairs that are in the same cluster under $\mathcal{D}$ but in different clusters under $\tilde{\mathcal{D}}$,
\item[$n_{10}$] is the number of point pairs that are in different clusters under $\mathcal{D}$ but in the same cluster under $\tilde{\mathcal{D}}$,
\item[$n_{11}$] is the number of point pairs that are in the same cluster under both $\mathcal{D}$ and $\tilde{\mathcal{D}}$.
\end{itemize}

\change{Tables~\ref{tab:proportion} and \ref{tab:Rand} show the proportions of selecting the correct number of clusters and the average Rank Index values
based on our method under various situations. As expected, our method performs better for a smaller $a$ and a larger $n$.}

\begin{table}[tb]\centering
\caption{Proportions of selecting the correct number (i.e., $K=2$) of clusters under various situations based on 500 simulated replicates.}
\begin{tabular}{c|cc|cc}
\hline 
$\alpha_2$ & \multicolumn{2}{c|}{$a=0.01$} & \multicolumn{2}{c}{$a=0.1$}\tabularnewline
\cline{2-5} & $n=100$ & $n=500$ & $n=100$ & $n=500$\tabularnewline
\hline 
$0.1$ & 0.326 & 0.296 & 0.286 & 0.266\tabularnewline
$0.2$ & 0.446 & 0.736 & 0.402 & 0.636\tabularnewline
$0.3$ & 0.560 & 0.800 & 0.474 & 0.724\tabularnewline
$0.4$ & 0.642 & 0.794 & 0.580 & 0.706\tabularnewline
$0.5$ & 0.694 & 0.834 & 0.648 & 0.660\tabularnewline
\hline 
\end{tabular}
\label{tab:proportion}
\end{table}

\begin{table}[tb]\centering
\caption{Average Rand index values based on the number of clusters selected by BIC under various situations based on 500 simulated replicates.}
\begin{tabular}{c|cc|cc}
\hline 
$\alpha_2$ & \multicolumn{2}{c|}{$a=0.01$} & \multicolumn{2}{c}{$a=0.1$}\tabularnewline
\cline{2-5} & $n=100$ & $n=500$ & $n=100$ & $n=500$\tabularnewline
\hline 
$0.1$ & 0.563 & 0.547 & 0.556 & 0.544\tabularnewline
$0.2$ & 0.603 & 0.746 & 0.582 & 0.663\tabularnewline
$0.3$ & 0.650 & 0.845 & 0.609 & 0.748\tabularnewline
$0.4$ & 0.689 & 0.884 & 0.660 & 0.778\tabularnewline
$0.5$ & 0.735 & 0.905 & 0.686 & 0.793\tabularnewline
\hline 
\end{tabular}
\label{tab:Rand}
\end{table}

\section{An application to precipitation data in Colorado}

In this section, we applied our method to a precipitation dataset in Colorado.
The dataset can be obtained from the Geophysical Statistics Project at the National Center for Atmospheric Research
(\url{http://www.image.ucar.edu/GSP/Data/US.monthly.met/CO.html}),
which has been analyzed previously by Paciorek and Schervish (2006) and Qadir \textit{et al.}~(2021).
It consists of monthly total precipitation (in mm) recorded at 367 weather stations across Colorado from 1895 to 1997.
It is well known that Western Colorado is mountainous with more significant topographical variability than Eastern Colorado.

\begin{figure}[tb]\centering
\begin{tabular}{cc}
\includegraphics[scale=0.38,trim={0cm 1cm 0.5cm 0cm},clip]{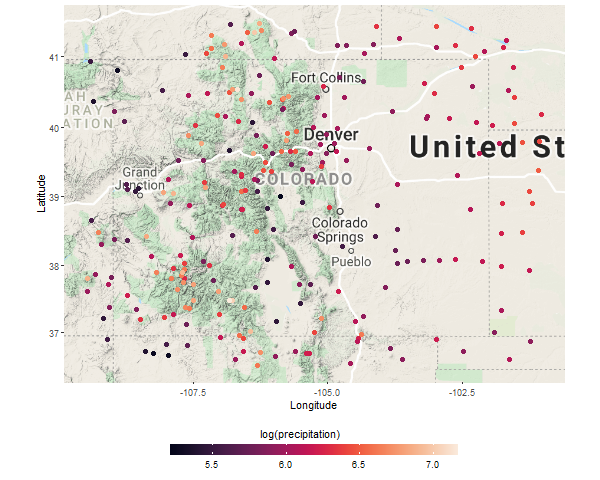} & \includegraphics[scale=0.38,trim={0cm 1cm 0.5cm 0cm},clip]{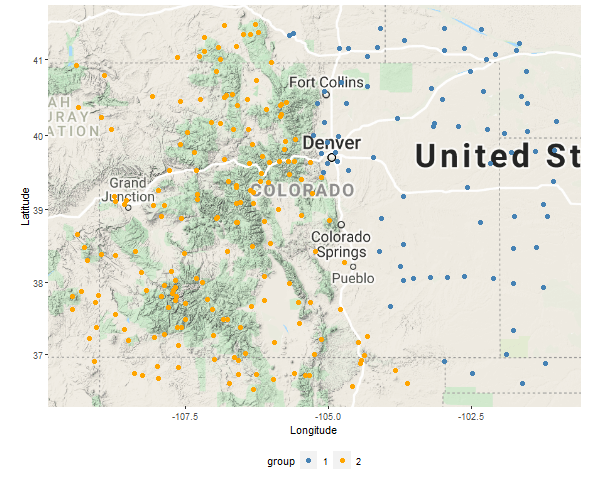}\\
(a) & (b)\\
\includegraphics[scale=0.38,trim={0cm 1cm 0.5cm 0cm},clip]{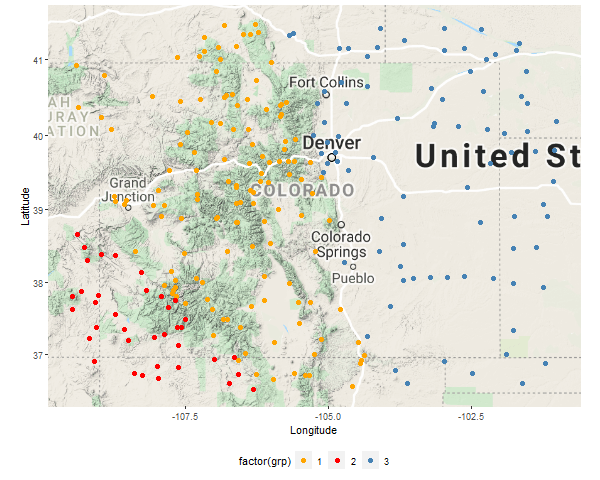} & \includegraphics[scale=0.38,trim={0cm 1cm 0.5cm 0cm},clip]{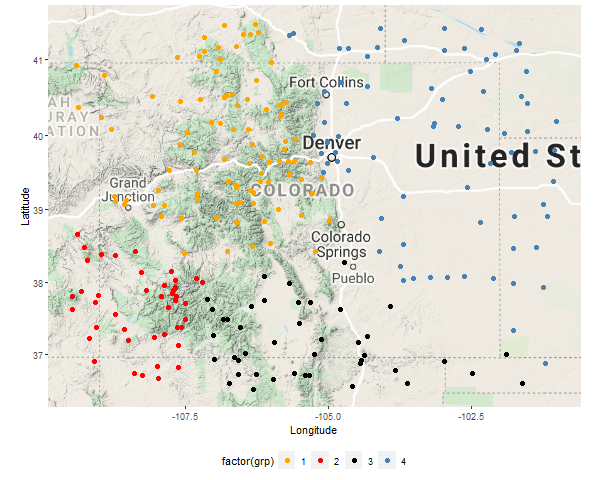}\\
(c) & (d)
\end{tabular}
\caption{(a) Precipitation amounts (mm in log scale) at 254 stations in Colorado in 1992;
(b) Two subregions obtained by the proposed methods;
(c) Three subregions obtained by the proposed methods;
(d) Four subregions obtained by the proposed methods.}
\label{fig:groups-for-test}
\end{figure}

Following Qadir \textit{et al}.~(2021), we considered the cumulative precipitations in the year 1992 and analyzed the data observed at 254 stations with no missing observations after applying the log transformation.
Figure \ref{fig:groups-for-test}(a) shows the precipitation data we analyzed.

\change{We estimated $\tau^2$ based on \eqref{eq:robust} and \eqref{eq:nugget} by selecting a small $d^*_1$ and $d^*_2$ so that $m_1=m_2=250$.}
Applying the proposed test of \eqref{eq:test} described in Section \ref{sect:test},
we obtained a p-value smaller than $0.01$ for testing spatial stationarity,
suggesting that the underlying process is likely nonstationary.
We then segmented the process into stationary processes based on subregions
by applying the proposed spatial segmentation \change{based on} \eqref{eq:partition} introduced in Section \ref{sec:Voronoi}.
From \eqref{eq:BIC}, we obtained the BIC values \change{$359.75$, $220.31$, $213.82$, and $228.96$},
for $K=1,\dots,4$, respectively, where $K=1$ corresponds to the stationary exponential model. The smallest BIC value is achieved at $K=3$.
Figure \ref{fig:groups-for-test}(b)-(d) shows the segmentation results based on $K=2,3,4$.
Even though we did not utilize any additional information (such as elevation) other than precipitations,
Colorado Eastern Plains, which tend to have a different climate pattern from the rest,
are automatically segmented as a subregion for $K\in\{2,3,4\}$,
demonstrating that the proposed spatial segmentation method is effective.

\change{We also investigated whether the proposed segmentation enhances spatial prediction.
We randomly split the data into training data $\{z(\bm{s}_i):i\in\mathcal{I}_{\mathrm{train}}\}$ (consisting of 204 observations) and test data $\{z(\bm{s}_i):i\in\mathcal{I}_{\mathrm{test}}\}$ (with 50 observations).
Using the training data, we applied the proposed spatial segmentation method \eqref{eq:partition} introduced in Section \ref{sec:Voronoi} with $K=1,\dots,4$.
Upon identifying the K subregions through our methodology, we conducted spatial prediction by fitting an exponential covariance model to each subregion independently,
operating under the assumption that the data were generated from \eqref{eq:data}.
Our approach considered y(·) as a piecewise stationary process, in line with the decomposition.
For every subregion, the model parameters were estimated using Maximum Likelihood (ML).
Subsequently, we harnessed ordinary kriging from equation \eqref{eq:ok} to derive the predictive surface for each subregion.
To gauge the performance of our predictors, we utilized the root mean squared prediction error (RMSPE) criterion:}
\[
\change{\mathrm{RMSPE}=\bigg\{\frac{1}{50}\sum_{i\in\mathcal{I}_{\mathrm{test}}}\big(\tilde{y}(\bm{s}_i)-z(\bm{s}_i)\big)^2\bigg\}^{1/2}.}
\]
\change{We also evaluated the performance of probabilistic forecast using the continuous ranked probability score (CRPS, Geniting and Raftery, 2007):}
\[
\change{\mathrm{crps}(F, z)= \int_{-\infty}^{\infty}\left(F(t) - I(t \geq z) \right)^2 \, \mathrm{d}t,}
\]
\change{where $F(\cdot)$ is the predictive cumulative distribution function, $z \in \mathbb{R}$ is an observation, and $I(\cdot)$ is an indicator function.
We computed the CRPS based on test data:}
\begin{equation*}
\change{\mathrm{CRPS} = \frac{1}{50}\sum_{i\in\mathcal{I}_{\mathrm{test}}} \mathrm{crps}\, (\tilde{F}(\bm{s}_i),z(\bm{s}_i)),}
\end{equation*}

\noindent \change{where for $i\in\mathcal{I}_{\mathrm{test}}$, $\tilde{F}(\bm{s}_i)$ is a generic predictive cumulative distribution function of $z(\bm{s}_i)$.}

\begin{figure}[tb]\centering
\begin{tabular}{cc}
\includegraphics[scale=0.28,trim={0cm 1cm 0cm 0cm},clip]{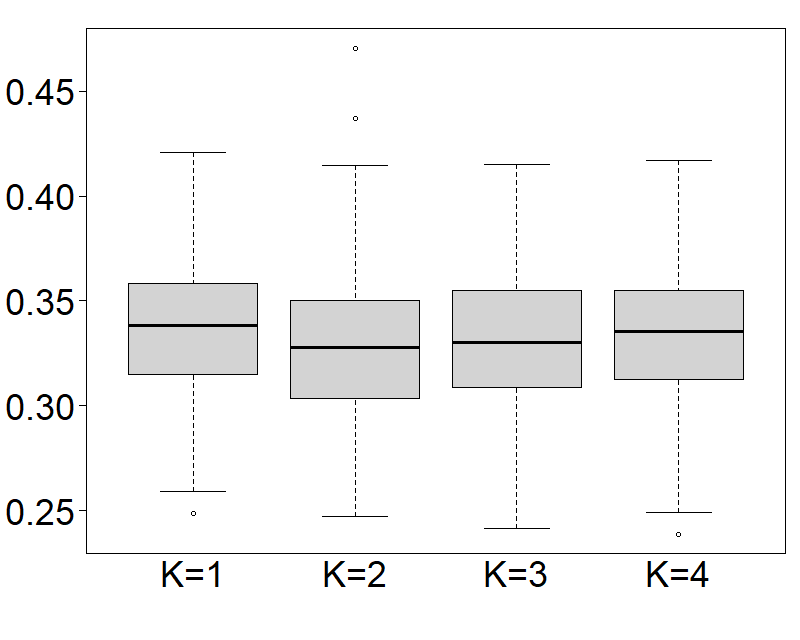} &
\includegraphics[scale=0.28,trim={0cm 1cm 0cm 0cm},clip]{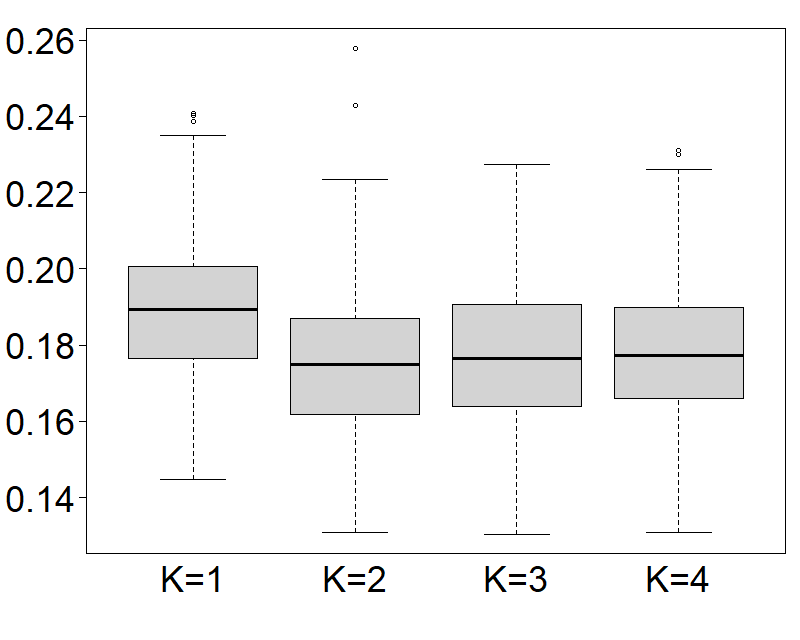}\\
(a) & (b)
\end{tabular}
\caption{\change{Prediction performances of the precipitation data in Colorado based on 200 pairs of randomly split training and test data:
 (a) Boxplots of RMSPEs; (b) Boxplots of CRPSs.}}
\label{fig:MSPE}
\end{figure}

\change{We randomly split the data into training and test data 200 times and obtained 200 predicted values and prediction standard deviations at each location.
Figure~\ref{fig:MSPE} shows boxplots of the RMSPE and CRPS values for $K=1,\dots,4$.
Our method performs better than the stationary model in terms of RMSPE and CRPS regardless of $K=2,3,4$.}

\section{Summary}

\change{We develop a statistic to track nonstationarity by focusing on a microergodic parameter. This innovation enables us to simultaneously detect changes in both spatial variances and spatial ranges, from which we can segment the region into stationary components using Voronoi tessellations. The proposed method is designed for data observed at irregularly spaced locations without repeated measurements.}

\change{Additionally, we introduce a novel test to detect the nonstationarity of a spatial process. Our test is not only computationally efficient, but it also properly controls the Type-I error rate, proving to be more powerful than existing methods. Compared to the test by Bandyopadhyay and Rao (1997), which tends to underperform with an irregular sampling design, our test remains largely unaffected by the irregularity of data locations.}

\change{The proposed stationarity test offers another advantage: it can point out where the nonstationarity occurs once rejected. As a result, we can perform kriging by applying a stationary model to each component separately. It is also conceivable to take this further by establishing a divide-and-conquer strategy to combine the results. These avenues present promising research directions, especially when dealing with massive spatial data. Further investigations along these lines, including the construction of nonstationary models based on locally stationary processes and the development of scalable methods for kriging, are of significant interest but fall beyond the scope of this paper. We intend to explore these areas in future work.}

\appendix

\section{}

\change{In this section, we display the distributions of p-values for various scenarios under $H_0$ in Section~\ref{sec:test}.}

\setcounter{figure}{0}
\begin{figure}\centering
\includegraphics[scale=0.39]{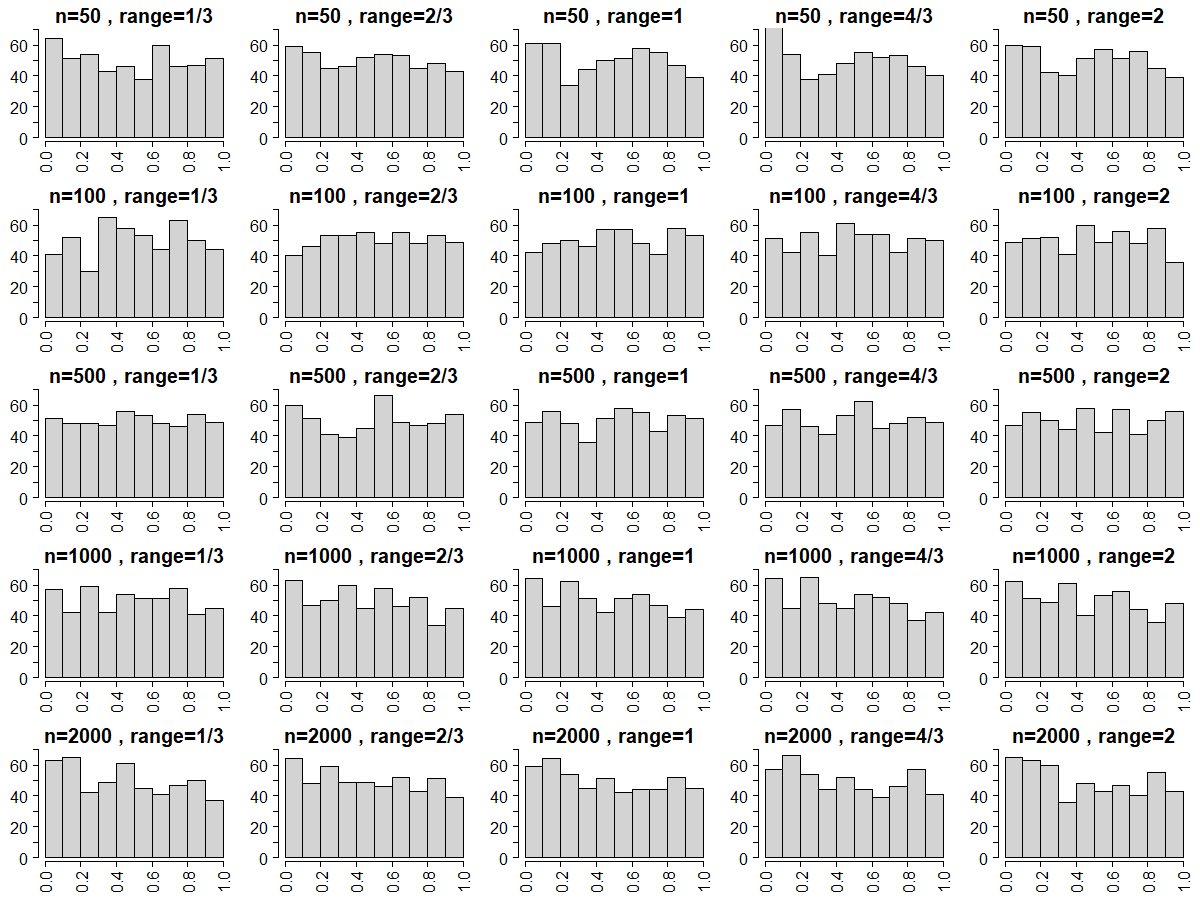}
\caption{The distributions of p-values under $H_0$ for various scenarios with $\change{\tau^2}=0$ under a uniform sampling design.}
\label{fig:pval-1a}
\end{figure}

\begin{figure}\centering
\includegraphics[scale=0.39]{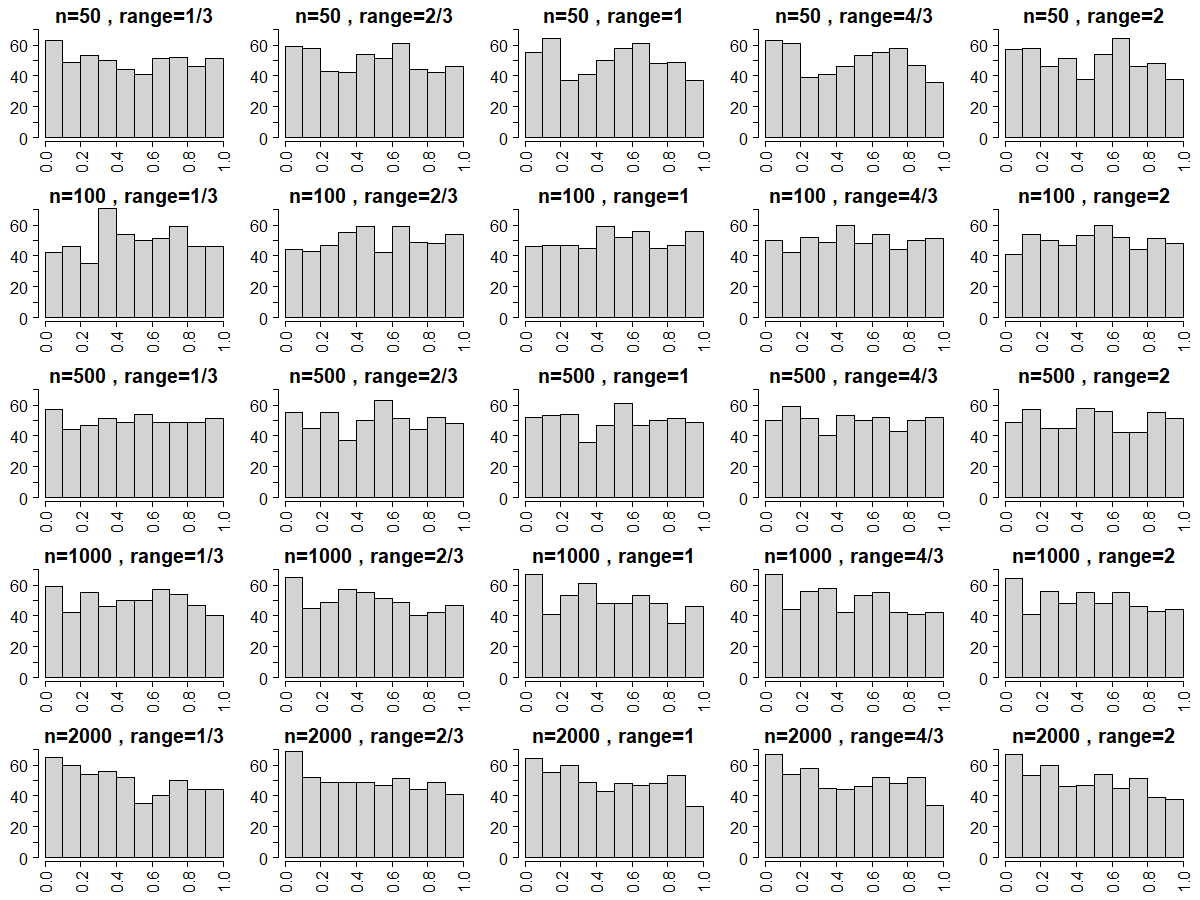}
\caption{The distributions of p-values under $H_0$ for various scenarios with $\change{\tau^2}=0.01$ under a uniform sampling design.}
\label{fig:pval-1b}
\end{figure}

\begin{figure}\centering
\includegraphics[scale=0.39]{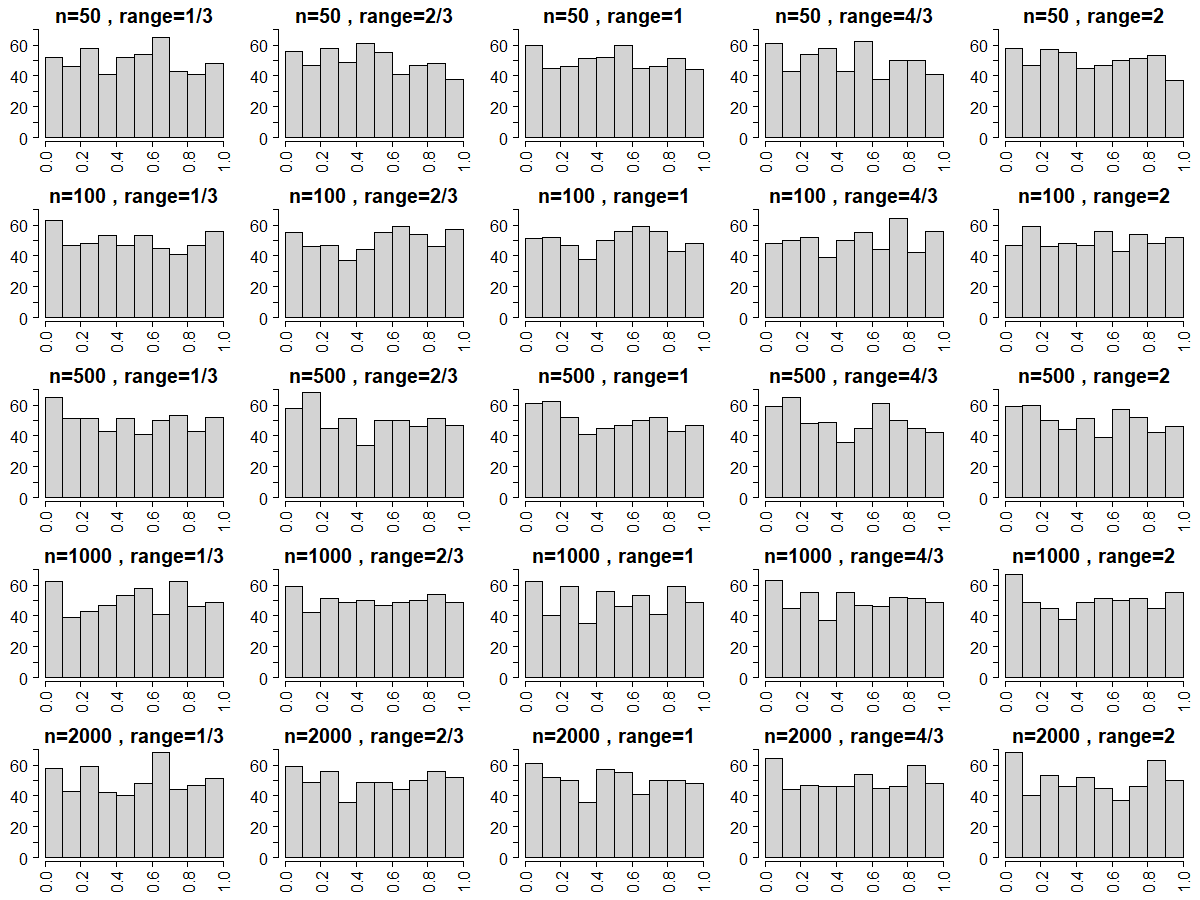}
\caption{The distributions of p-values under $H_0$ for various scenarios with $\change{\tau^2}=0$ under a clustered sampling design.}
\label{fig:pval-2a}
\end{figure}

\begin{figure}\centering
\includegraphics[scale=0.39]{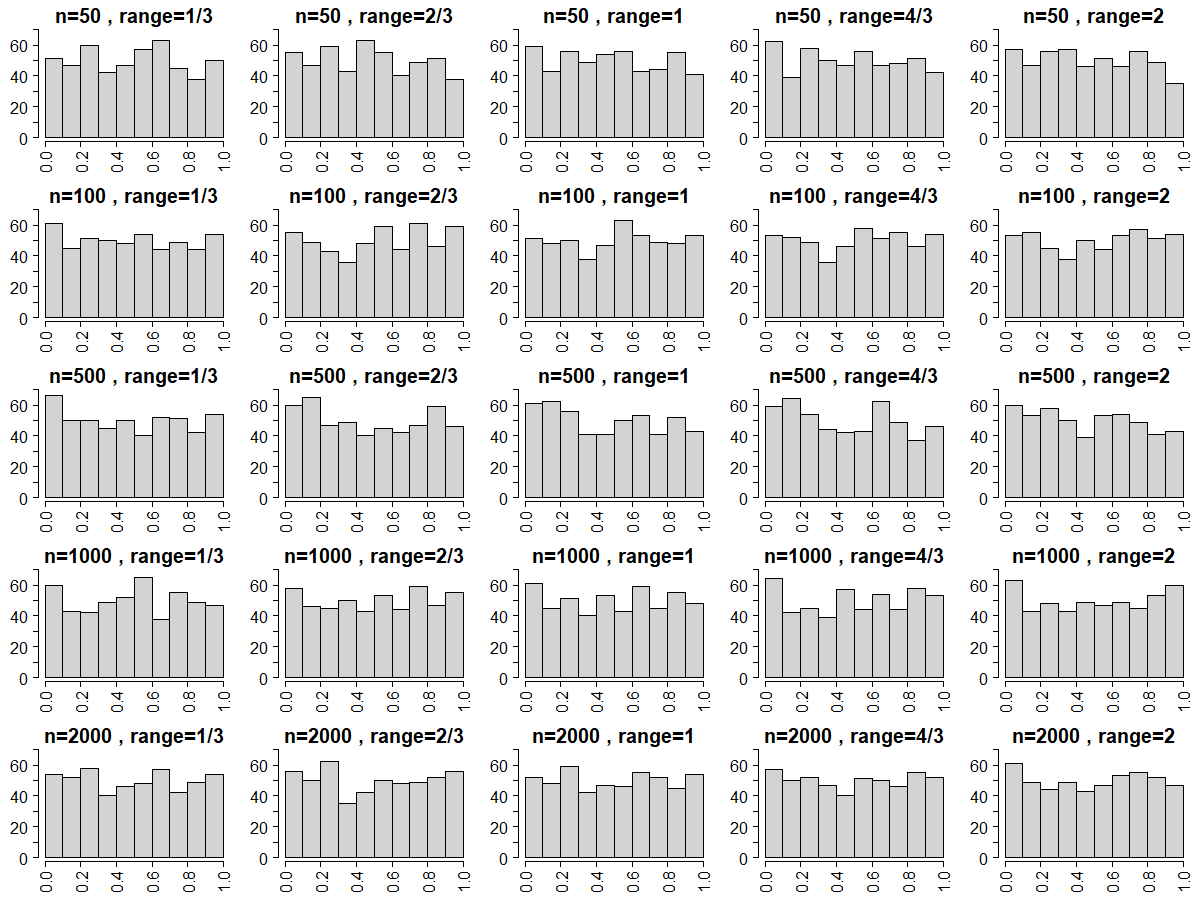}
\caption{The distributions of p-values under $H_0$ for various scenarios with $\change{\tau^2}=0.01$ under a clustered sampling design.}
\label{fig:pval-2b}
\end{figure}

\section*{References}

\begin{description}
\item \change{Anselin, L.~(1995). Local indicators of spatial association--LISA, \textit{Geographical Analysis}, \textbf{27}, 93--115.}
\item Bandyopadhyay, S. and Rao, S.~S.~(2017). A test for stationarity for irregularly spaced spatial data,
  \textit{Journal of Royal Statistical Society, Series B}, \textbf{79}, 95--123.
\item Cressie, N.~(1993). Statistics for Spatial Data, rev.~edn, Wiley, New York, NY.
\item Cressie, N.~and Hawkins, D.~M.~(1980). Robust estimation of the variogram: I, \emph{Mathematical Geology}, \textbf{12}, 115--125. 
\item Fuentes, M.~(2005). A formal test for non-stationarity of spatial stochastic processes. \textit{Journal of Multivariate Analysis},
  \textbf{96}, 30--54. 
\item \change{Gneiting, T.~and Raftery, A.~E.~(2007). Strictly proper scoring rules, prediction, and estimation,
  \textit{Journal of the American Statistical Association}, \textbf{102}, 359--378.}
\item \change{Guinness, J., and Fuentes, M.~(2015).~Likelihood approximations for big nonstationary spatial temporal lattice data,
  \textit{Statistica Sinica}, \textbf{25}, 329--349.}
\item Jun, M.~and Genton, M.~(2012) A test for stationarity of spatio-temporal random fields on planar and spherical domains,
  \textit{Statistica Sinica}, \textbf{22}, 1737--1764.
\item Mat\'{e}rn, B.~(1986). \textit{Spatial Variation}, 2nd ed., Springer-Verlag, Berlin.
\item \change{Muyskens, A., Guinness, J., and Fuentes, M.~(2022). Partition-based nonstationary covariance estimation using the stochastic score approximation,
  \textit{Journal of Computational and Graphical Statistics}, \textbf{31}, 1025--1036.}
\item Nidheesh, N., Nazeer, K. A., and Ameer, P. M. (2017). An enhanced deterministic K-Means clustering algorithm for cancer subtype prediction from gene expression data. \emph{Computers in biology and medicine}, \textbf{91}, 213-221.
\item Paciorek, C. J. and Schervish, M. J. (2006). Spatial modelling using a new class of nonstationary covariance functions.
  \textit{Environmetrics}, \textbf{17}, 483-506.
\item Qadir, G. A., Sun, Y. and Kurtek, S. (2021). Estimation of spatial deformation for nonstationary processes via variogram alignment. 
  \textit{Technometrics}, \textbf{63}, 548--561.
\item Rand, W.~M.~(1971). Objective criteria for the evaluation of clustering methods,
  \textit{Journal of the American Statistical Association}, \textbf{66}, 846--850.
\item Schwarz, G.~(1978). Estimating the dimension of a model, \textit{The Annals of Statistics}, \textbf{6}, 461--464.
\item \change{Tibshirani, R., Saunders, M., Rosset, S., Zhu, J., and Knight, K.~(2005).
  Sparsity and smoothness via the fused lasso, \textit{Journal of the Royal Statistical Society, Series B}, \textbf{67}, 91--108.}
\item Voronoi, G. (1908). Nouvelles applications des paramètres continus à la théorie des formes quadratiques.
  Premier mémoire. Sur quelques propriétés des formes quadratiques positives parfaites,
  \textit{Journal f\"{u}r die reine und angewandte Mathematik (Crelles Journal)}, \textbf{1908}, 97--102.
\item \change{Zhang, H.~(2004).~Inconsistent estimation and asymptotically equal interpolations in model-based geostatistics,
  \textit{Journal of the American Statistical Association}, \textbf{99}, 250--261.}
\end{description}

\end{document}